\documentclass[iop]{emulateapj}
\usepackage{natbib}
\usepackage{mathtools}
\usepackage{hyperref}

\def\statdla{4}
\def\statdz{123.3}
\def\statdx{229.7}
\def\totqso{878}
\def\nqso{767}
\def\nstat{463}

\def\hi{H\,{\sc i}}
\def\fn{$f(N_{\rm{H}\textsc{i}},X)$}
\def\nh{$N_{\rm{H}\textsc{i}}$}
\def\kms{km~s$^{-1}$}
\def\lognh{$\log N_{\rm{H}\textsc{i}}$}
\def\lox{$\ell_{\rm{DLA}}(X)$}
\def\rhohi{$\rho_{\rm{H}\textsc{i}}^{\rm{DLA}}$}

\shorttitle{The {\hi} content of the Universe over the past 10 Gyrs}
\shortauthors{Neeleman et al.}

\begin{document}

\title{The {\hi} content of the Universe over the past 10 Gyrs}

\author{Marcel Neeleman\altaffilmark{1,2,3}, J. Xavier Prochaska\altaffilmark{2}, Joseph Ribaudo\altaffilmark{4},
Nicolas Lehner\altaffilmark{5},J. Christopher Howk\altaffilmark{5},Marc Rafelski\altaffilmark{6}, 
Nissim Kanekar\altaffilmark{7}}
\altaffiltext{1}{Department of Physics and Center for Astrophysics and Space Sciences, UCSD, La Jolla, CA 92093, USA}
\altaffiltext{2}{Department of Astronomy \& Astrophysics, UCO/Lick Observatory, 1156 High Street, 
University of California, Santa Cruz, CA 95064, USA}
\altaffiltext{3}{marcel@ucsc.edu}
\altaffiltext{4}{Department of Physics, Utica College, 1600 Burrstone Road, Utica, NY 13502}
\altaffiltext{5}{Department of Physics, University of Notre Dame, Notre Dame, IN 46556}
\altaffiltext{6}{NASA Postdoctoral Program Fellow, Goddard Space Flight Center, Code 665, Greenbelt, MD 20771, USA}
\altaffiltext{7}{National Centre for Radio Astrophysics, Tata Institute of Fundamental Research, 
Pune University, Pune 411007, India}

\begin{abstract} 

We use the Hubble Space Telescope (HST) archive of ultraviolet (UV)
quasar spectroscopy to conduct the first blind survey for damped Ly-$\alpha$ 
absorbers (DLAs) at low redshift ($z < 1.6$).  Our statistical sample includes 
{\nstat} quasars with spectral coverage spanning a total redshift path 
$\Delta z = \statdz$ or an
absorption path $\Delta X = \statdx$. 
Within this survey path, we identify {\statdla} DLAs defined as
absorbers with {\hi} column density {\nh} $\ge 10^{20.3}$\,cm$^{-2}$,
which implies an incidence per absorption length
{\lox} $= 0.017^{+0.014}_{-0.008}$ at a median survey path redshift of $z=0.623$. While
our estimate of {\lox} is lower than earlier estimates at $z \approx 0$ from {\hi}~21\,cm emission
studies, the results are consistent within the measurement uncertainties. Our dataset is 
too small to properly sample the {\nh} frequency distribution function {\fn}, but the observed 
distribution agrees with previous estimates at $z > 2$. Adopting the $z > 2$ shape of {\fn}, 
we infer an {\hi} mass density at $z \sim 0.6$ of 
{\rhohi} $= 0.25^{+0.20}_{-0.12} \times 10^8 M_{\odot} \rm{Mpc}^{-3}$.  
This is significantly lower than previous estimates from targeted DLA surveys with 
the HST, but consistent with results from low-$z$ {\hi}~21\,cm observations, and suggests that 
the neutral gas density of the universe has been decreasing over the past 10 Gyrs.

\end{abstract}


\section{Introduction}
\label{sec:intro}

Galaxy formation and evolution are critically dependent on the gas within and surrounding a galaxy. 
As galaxies evolve, gas is accreted onto the galaxy and expelled through various processes such 
as activity of an active galactic nucleus and stellar feedback. Providing observational constraints on 
this gas is therefore paramount in understanding how galaxies formed and evolved. At very low redshifts, 
neutral gas has been studied in detail using the {\hi}~21\,cm line. Unfortunately, such observations are 
limited with current facilities to very low redshifts, $z \lesssim 0.25$ 
\citep[e.g.,][]{Zwaan2001,Catinella2008,Fernandez2013,Catinella2015}. 

\begin{deluxetable*}{llllcllclll}
\tabletypesize{\scriptsize}
\tablecaption{{QSO sample}
\label{tab:qso}}
\tablewidth{0pt}
\tablehead{
\colhead{QSO} &
\colhead{$z_{\rm{em}}$\,\tablenotemark{a}} &
\colhead{Instrument} &
\colhead{Grating} &
\multicolumn{3}{c}{Search Path} &
\multicolumn{3}{c}{Statistical Path} &
\colhead{Proposal ID} \\
\cline{5-7} &
\cline{7-9} 
\colhead{} &
\colhead{} &
\colhead{} &
\colhead{} &
\colhead{F$_{\rm{search}}$\,\tablenotemark{b}} &
\colhead{Min $z$} &
\colhead{Max $z$} &
\colhead{F$_{\rm{stat}}$\,\tablenotemark{c}} &
\colhead{Min $z$} &
\colhead{Max $z$} &
\colhead{}
}
\startdata
J0000$-$1245 & 0.200 & COS & G130M-G160M & 1 & 0.010 & 0.299 & 1 & 0.010 & 0.188 & 12604\\
J0001$+$0709 & 3.234 & STIS & G230L & 1 & --- & --- & 0 & --- & --- & 8569\\
J0004$-$4157 & 2.760 & FOS & G270H & 1 & 1.501 & 1.695 & 0 & --- & --- & 6577\\
J0005$+$0524 & 1.900 & FOS & G270H-G190H & 1 & 0.829 & 1.695 & 1 & 0.829 & 1.695 & 4581,6705\\
J0005$-$5006 & 0.033 & COS & G130M-G160M & 1 & 0.010 & 0.133 & 1 & 0.010 & 0.022 & 12936
\enddata
\tablenotetext{a}{Emission redshift of quasar.}
\tablenotetext{b}{Search flag: (0) Low S/N or bad spectrum; (1) Included; (2) BAL quasar.}
\tablenotetext{c}{Statistical flag: (0) Non-Statistical; (1) Statistical; (2) Galaxy Sample.}
\tablecomments{(Omitted from this portion of the table for brevity are
  the columns for alternate QSO name, right ascension and
  declination. These columns and the table in its entirety is
  available as a machine-readable table in the online journal. The
  complete sample table is also listed at the end of this manuscript
  minus the aforementioned columns.)}
\end{deluxetable*}

To study how the mass and distribution of neutral gas has evolved over cosmic time, we 
need to measure the cosmic density of this gas over a large redshift
range. Such a study can be done by studying the gas through Ly-$\alpha$ absorption in 
quasar spectra \citep{Wolfe1986}. Previous studies have shown that the highest column density
absorbers, the damped Ly-$\alpha$ systems (DLAs), which have neutral
hydrogen column densities of {\nh}~$\ge 10^{20.3}$\,cm$^{-2}$, contain
the bulk of the neutral gas at both low and high redshifts
\citep[e.g.,][]{Prochaska2005,Wolfe2005,Zwaan2005,OMeara2007}. Observations of DLAs 
therefore provide an excellent opportunity to constrain the neutral hydrogen content 
of our Universe over a large redshift range. 

To quantify the overall {\hi} content of our Universe, we use a
single quantity known as the neutral hydrogen column density
distribution function, {\fn}, \citep[e.g.,][]{Tytler1987,Lanzetta1991,
Prochaska2005}. {\fn} is a quantitative
description of the large scale neutral hydrogen distribution of our
Universe, and its zeroth and first moment yield the line density
of DLAs, {\lox}, and the total hydrogen mass density of the Universe,
{\rhohi}. 

At high redshifts, $z \gtrsim 2$, {\fn} for DLAs has been measured by many 
authors (\citealp[e.g.,][]{Prochaska2009,Noterdaeme2012}; but see 
\citealp{Crighton2015} and \citealp{Sanchez2015}), using large optical
surveys such as the Sloan Digital Sky Survey \citep[SDSS;][]{Abazajian2009}. 
These studies find that the shape of {\fn} is invariant between $z \sim 2$ and $z \sim 3.5$, while the 
normalization is observed to evolve, decreasing by a factor of $\approx 1.3-2$. 
This implies a concomitant decrease in {\lox} and {\rhohi}. Such evolution may 
be explained by either gas conversion into stars over this redshift range or 
feedback processes that expel gas from galaxies \citep[e.g.,][]{Prochaska2009}. 

A key feature of {\fn} and its moments is that they appear to converge by $z \sim
2$ to the present-day values estimated using {\hi}~21\,cm studies 
\citep{Zwaan2005,Braun2012,Delhaize2013,Hoppmann2015}. This suggests that these 
quantities have remained essentially unchanged over the last 10 billion years of 
galaxy evolution. Unfortunately, at $z \lesssim 2$, it is difficult to measure {\fn}  
directly, because the effectiveness of optical surveys plummets due to the atmospheric
absorption of ultraviolet (UV) radiation \citep{Lanzetta1995}. This problem can be 
circumvented by observing from space with the UV spectrographs on the Hubble Space 
Telescope (HST). However, due to the expense of these observations and the scarcity 
of DLAs in random lines of sight, large-scale blind surveys comparable to the SDSS have 
not been feasible within the limited time allocations of single observing programs. 

To increase the rate of detection of high {\hi} column density absorbers
along quasar sightlines, previous studies at $z < 2$ have used several 
types of pre-selection methods, with the most common approach
being the use of strong Mg\,{\sc ii} absorption to pre-select DLA and 
{\hi}~21\,cm candidates \citep[e.g.,][]{Briggs1983,Rao2000,Rao2006,Kanekar2009}. 
At redshifts $z \gtrsim 0.1$, the Mg\,{\sc ii} doublet is shifted
into the optical regime. As the vast majority of DLAs show strong
Mg\,{\sc ii} absorption (rest equivalent width, $W^{\lambda 2796}_0 >
0.5$\,{\AA}), pre-selecting quasars with strong Mg\,{\sc ii} absorption
will significantly increase the detection rate of DLAs in the sample. However,
it has not so far been straightforward to understand the biases in 
such pre-selections, which are critical to obtain accurate estimates of {\fn} 
for DLAs \citep[e.g.,][]{Rao2006}. 

Fortunately, 20 years of HST observations with a variety of UV spectrographs have 
resulted in a large sample of observed quasars. As HST nears the end of its mission, 
the time has come to explore this large data set, and perform a study in the UV to 
evaluate {\fn}, {\lox} and {\rhohi} at $z < 1.6$, similar to earlier studies performed 
in the optical regime. In this paper, we use the HST spectroscopic archive to obtain 
a measurement of the above quantities between $z \sim 0.01$ and $z \sim 1.6$, covering the 
past 10 billion years of the Universe. The sample for this study is described in 
Section~\ref{sec:sample}, and the method is described in Section~\ref{sec:method}. Our 
results are presented in Section~\ref{sec:results} and discussed in Section~\ref{sec:disc}. 
Throughout this paper we adopt an $(\Omega_{\rm{M}}, \Omega_{\Lambda},h) =
(0.3,0.7,0.7)$ cosmology. 

\begin{figure*}[!t]
\epsscale{1.17}
\plottwo{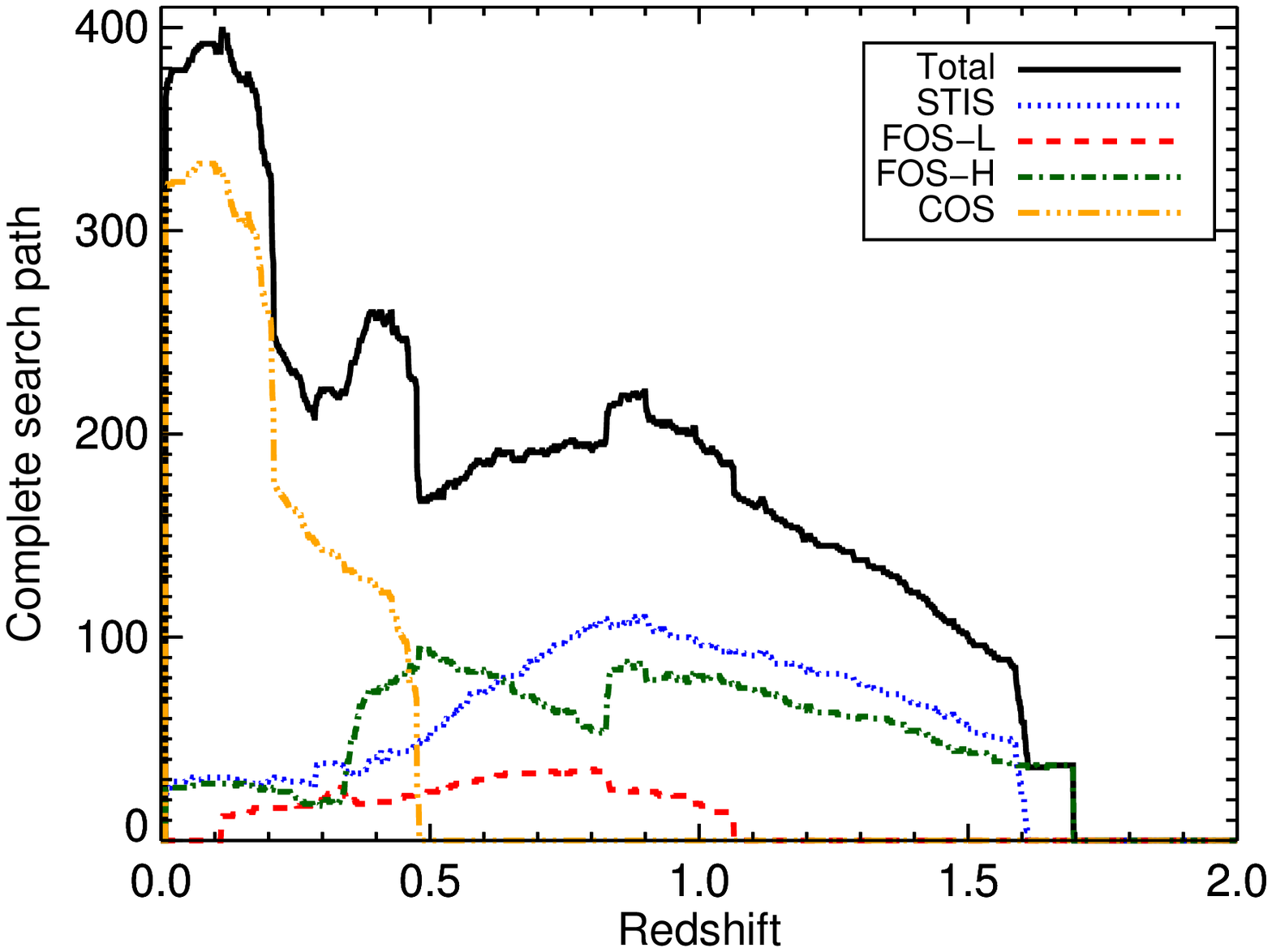}{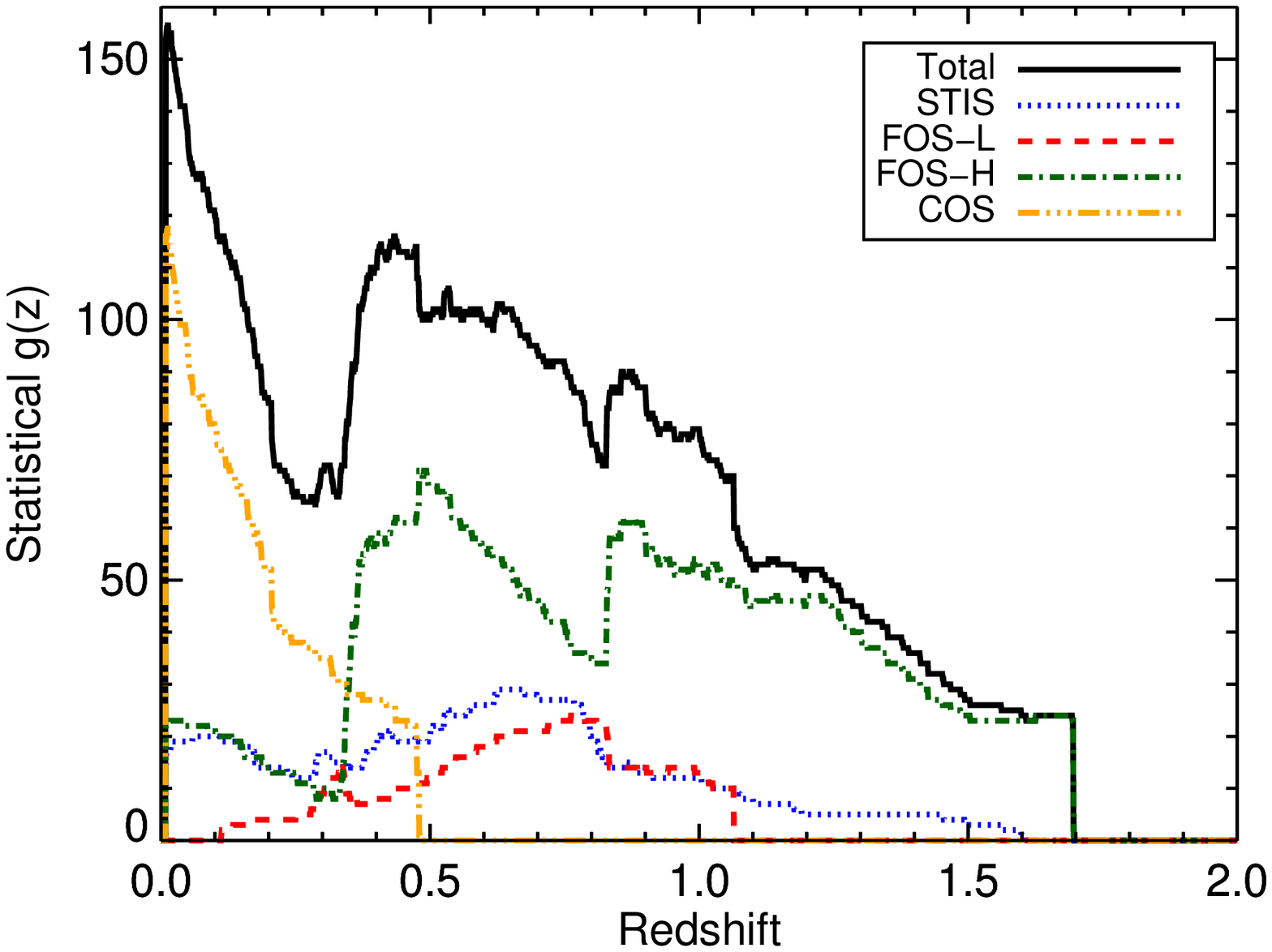}
\caption{\emph{Left panel:} The complete search path for our target sample of $\nqso$ quasars. 
  The search paths of the individual spectrographs as well as the combined search path are
  plotted. \emph{Right panel:} The statistical search path $g(z)$ for the statistical
  sample of quasars (i.e., those quasars with f$_{\rm{stat}} = 1$). The total number of 
  quasars included in this statistical sample is {\nstat}.}
\label{fig:gz}
\end{figure*} 

\section{Sample Selection}
\label{sec:sample}

To measure the amount of neutral hydrogen between $z \sim 0.01$ and $z
\sim 1.6$, we have assembled a large sample of quasars observed with
medium resolution spectrographs on the Hubble Space
Telescope. Specifically, we performed a search of the HST archive for quasars
observed with either the Space Telescope Imaging Spectrograph (STIS),
the Faint Object Spectrograph (FOS) or the Cosmic Origins Spectrograph
(COS). These instruments have gratings that provide enough spectral resolution for a high
fidelity search for strong absorption line systems (see Table \ref{tab:spec} and Section
\ref{sec:res}). We did not include the Goddard High Resolution
Spectrograph (GHRS), as its spectral coverage is too small to provide a
meaningful search path. In total, a sample of {\totqso} quasars were 
observed with at least one of these instruments and gratings.  

\begin{deluxetable}{llll}
\tabletypesize{\scriptsize}
\tablecaption{{Spectrographs}
\label{tab:spec}}
\tablewidth{0pt}
\tablehead{
\colhead{Spectrograph} &
\colhead{Grating} &
\colhead{Resolution\,\tablenotemark{a}} &
\colhead{Spectral Range\,\tablenotemark{b}} \\
\colhead{} &
\colhead{} &
\colhead{} &
\colhead{(\AA)}
}
\startdata
COS & G130M & 18000 & 1138 - 1466 \\
    & G160M & 18000 & 1410 - 1774 \\
FOS & G160L & 300 & 1351 - 2511 \\
 & G130H & 1400 & 1140 - 1606 \\
 & G190H & 1400 & 1590 - 2311 \\
 & G270H & 1400 & 2223 - 3277 \\
STIS & G140L & 1200 & 1120 - 1716 \\
 & G230L & 700 & 1573 - 3158
\enddata
\tablenotetext{a}{The median resolving power at the central wavelength.}
\tablenotetext{b}{The median central wavelengths for the complete data set.}
\end{deluxetable}

We will not outline the procedure here used to analyze the spectra. Instead, we refer the 
reader to the following papers that describe the reduction process in detail for each 
of the different instruments and gratings. The FOS~G160L and STIS spectra reductions are described 
in \citet{Ribaudo2011}. We note that the FOS~G160L data were cut off at a wavelength of 1350~\AA. 
This arbitrary cut-off does not affect the results in this paper (see Section \ref{sec:results}). 
The reduction of the higher resolution FOS spectra is detailed in \citet{Bechtold2002}.
Finally, the analysis of the COS spectra is described in \citet{Thom2011} and \citet{Meiring2011}. 

The reduced spectra were compiled into a single list, and, for quasars with multiple observations 
with different instruments, the spectra were combined into a single spectrum. In cases of 
overlapping spectral coverage, the higher resolution spectrum was used. We visually confirmed 
the emission redshift, and the quality of the spectra for all the quasars. Several quasars 
are not included either due to too low S/N or bad spectra ($99$ quasars). We also 
exclude any quasars that exhibit strong broad absorption line (BAL) features 
($12$ quasars). A total of {\nqso} quasars satisfy the above criteria and form the sample 
of this paper, as tabulated in Table \ref{tab:qso}.

\subsection{Statistical Sample} 
\label{sec:statsample}

To provide an accurate measurement of the {\hi} column density distribution function, 
it is critical to provide an unbiased quasar sample, as the inclusion of quasar sightlines 
targeted to study known absorption systems would bias {\fn} to higher values relative to the 
cosmic mean. Similarly, the inclusion of quasar sightlines that were targeted to contain a 
known absence of absorption systems would bias {\fn} low. We have therefore carefully considered 
the stated selection criteria for each of the observed quasars. In Table \ref{tab:qso}, we 
have included a flag that indicates if the target quasar was observed for a bias with respect to 
the presence of an absorber along the sightline.

The statistical flag (f$_{\rm{stat}}$) can take on three values. A
flag of 0 indicates the observed quasar was targeted because it contained either an
absorber (i.e., a known DLA or an Mg\,{\sc ii} system) or a lack of these systems
along the line of sight. A flag of 1 indicates that the quasar was
targeted independent of any known features along the line of
sight. Finally a flag of 2 indicates that the quasar sightline crosses
close to a previously recorded galaxy seen in emission. The true
statistical sample defined in this paper contains only the targets with a statistical flag 
equal to 1, consisting of a total of {\nstat} quasars. We also define an expanded sample 
in this paper which contains quasars with both  f$_{\rm{stat}}$ = 1 and f$_{\rm{stat}}$ = 2
($677$ quasars).

\section{Method}
\label{sec:method}

To search for absorption systems in the HST spectra, we apply a method similar 
to that described in \citet{Prochaska2005}, but slightly adapted for our lower 
redshift sample. Specifically, we define the search path for each quasar sightline, 
with the lower limit of the search path set by the signal-to-noise ratio (S/N) of the
spectrum. We determine the S/N by calculating a running median with a 20-pixel width 
for each wavelength. The lower limit to the search path is set to the wavelength for 
which this median S/N is greater than the minimum allowed S/N, which we take to be 4, 
or the wavelength of Ly-$\alpha$ at $z = 0$ ($\lambda_{\rm{Ly\alpha}} =
1215.6701$\,{\AA}), whichever is greater. The choice of 4 for this minimum S/N
is explained in Section~\ref{sec:S/N}. The upper limit to the search path is set to 
be either the end of the spectrum or the value $(1+z_{\rm{em}}+\rm{offset}) 
\lambda_{\rm{Ly\alpha}}$, whichever is smaller. Here, $z_{\rm{em}}$ is the quasar redshift;
the redshift offset is added to allow absorption systems to have slightly higher redshifts 
than that of the quasar. This offset redshift (chosen to be 0.1) also encompasses the 
uncertainty in the reported quasar emission redshifts. The complete search path is 
shown in the left panel of Figure~\ref{fig:gz}. 

For those quasars that are part of the statistical sample,
we also define a statistical search path, $g(z)$. The lower limit to
the statistical path is set again by the wavelength where the median
S/N exceeds the minimum S/N value of 4, or a wavelength of
$(1+\rm{offset}) \lambda_{\rm{Ly\alpha}}$, whichever is
greater. This time, we adopt a redshift offset of 0.01 which is $\sim$ 
3000\,{\kms} from our Galaxy. This offset is introduced to prevent any
biasing due to clustering in the Milky Way neighborhood. The upper limit is
set to the end of the spectrum or the value $\sqrt{(c-\Delta 
v)/(c+\Delta v)} (1+z_{\rm{em}}) \lambda_{\rm{Ly\alpha}}$, 
whichever is greater. Here, $\Delta v$ is taken to be 3000\,{\kms} which 
prevents clustering around the quasar from affecting our measurements. 
The search path for the statistical sample (f$_{\rm{stat}}$ = 1) is
shown in the right panel of Figure \ref{fig:gz}.  

After defining the search path, we run our search algorithm to find
candidate absorption systems along the search path to each quasar. The algorithm
searches regions of the spectrum that fall below a specified S/N cut,
which occurs in the absorption trough of strong absorbers. Specifically, 
the algorithm assigns to each pixel a DLA score, which is a
measure of how many pixels in a 3\,{\AA} window centered around this
pixel fall below the assigned S/N cut per pixel. We take a 3\,{\AA} window
because this is the core width of a {\nh} $= 10^{20.3}$ cm$^{-2}$ absorber
at $z=0.01$, while the S/N threshold is taken to be 2. The central pixels for
which greater than 60\,\% of the surrounding pixels in the 3\,{\AA} window fall 
below the S/N threshold are flagged by the algorithm as candidate absorption systems. 
Altogether, we recorded 139 such candidates from the complete sample.

The final step in the search for absorption systems is the visual follow-up of these 
candidates. We fit individual absorption systems using a custom IDL Voigt-profile fitting 
program for DLAs, $\tt x\_fitdla$, which is part of the the publicly available IDL 
library, XIDL\footnote{\url{http://www.ucolick.org/~xavier/IDL/}}. With this program,
we are able to simultaneously fit both the Voigt profile and the continuum of the 
quasar as described in detail in \citet{Prochaska2003a}. The largest source of uncertainty 
stems from the continuum placement.

As this process has inherently a human aspect to it, two of the authors (MN and JXP) 
independently carried out the fits to the candidate absorbers. The results were compared 
and, for most of the systems ($>$ 90\,\%), the measured {\hi} column density from both 
authors fell within the estimated 1-$\sigma$ uncertainty of the measurement. We therefore 
believe that our column density measurements are robust.

\begin{figure}[!t]
\epsscale{1.2}
\plotone{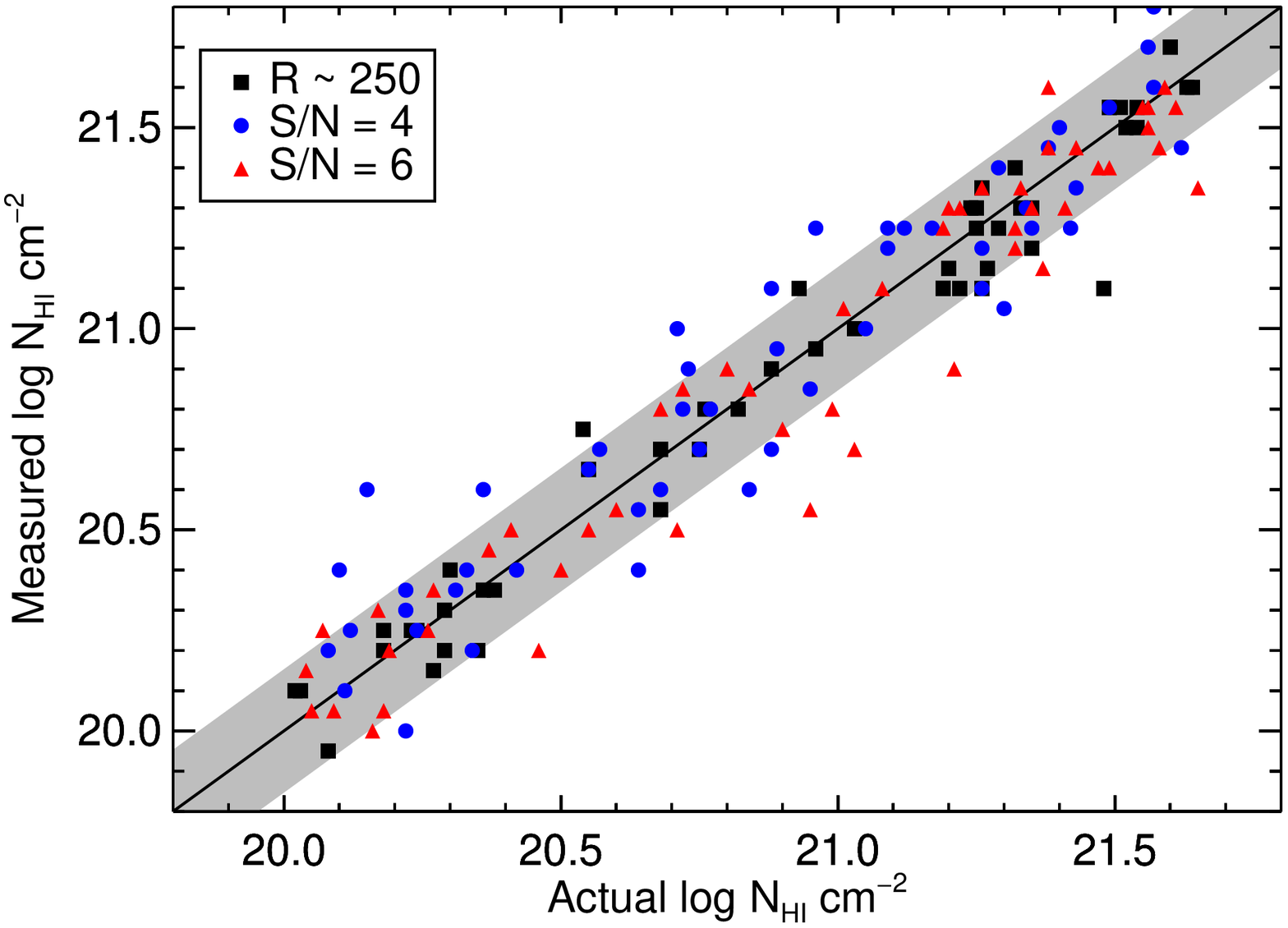}
\caption{Results from various mock tests to quantify the effect of
  resolution and S/N on our ability to accurately determine the {\hi} column
  density of individual absorbers, with the measured {\hi} column density for 
  simulated DLAs plotted versus their actual {\hi} column density. The solid line
  indicates the case where the measured and actual {\hi} column densities are 
  equal, while the gray area marks the 0.15\,dex range in dispersion around the actual 
  value (the mean uncertainty that we have assigned to our fits). The fact that the measured 
  {\hi} column densities cluster around the solid line, with no systematic shifts, indicate 
  that it is possible to obtain an accurate estimate of the {\hi} column density for both 
  low S/N ($\gtrsim 4$) and low resolution (R $\sim$ 250) data. }
\label{fig:fitdla}
\end{figure} 

 \subsection{Resolution Considerations}
 \label{sec:res}

Low resolution spectra could result in inaccurate {\hi} column density measurements and/or 
 non-detection of DLAs by the search algorithm. Our lowest resolution spectra are taken with 
 the low resolution gratings of the STIS~G230L 
 and FOS~G160L, which have resolutions as low as R~$\approx 250$. For such resolutions, only 
 about half of a resolution element falls within the DLA trough of a low column density 
 system, and therefore small deviations could result in a non-detection of such a DLA by 
 the search algorithm.
 
To assess the impact of resolution on the recovery process, we created
simulated spectra with similar S/N and resolution to the observational
data. We added artificial DLAs to the spectra with a range of
column densities. For this mock data set, the algorithm was successful in
recovering greater than 99\,\% of all of the absorbers with a column
density above $10^{20}$\,cm$^{-2}$. We conclude that the resolution of the
spectra is sufficient to accurately recover DLAs.

We also test our ability to recover correct {\hi} column density measurements by 
adding DLAs to our lowest resolution spectra and fitting these DLAs with the
fitting procedure described in Section~\ref{sec:method}. In particular, we added 50 DLAs 
with varying {\hi} column densities to our lowest resolution spectra (i.e. both STIS and FOS-L) 
and measured their {\hi} column densities via our fitting method. The results are shown 
in Figure~\ref{fig:fitdla}. As can be seen from the figure, we are able to accurately recover 
the actual {\hi} column density over the full range of input values, including the lowest
column density systems. Furthermore, the lower resolution 
does not cause any systematic under- or over-estimate of the {\hi} column density. The deviation 
around the actual value for this sample is 0.10\,dex, which is comparable to the 
fit uncertainties (which have a mean of 0.15\,dex) that we have assigned to the sample.

As a final test, we have reanalyzed the results in this paper with the lowest resolution
spectra omitted from the sample, and find no significant difference in any of the result presented.

\begin{deluxetable*}{llcllccl}
\tabletypesize{\scriptsize}
\tablecaption{{DLA sample}
\label{tab:dla}}
\tablewidth{0pt}
\tablehead{
\colhead{QSO Name} &
\colhead{Alternate QSO Name} &
\colhead{F$_{\rm{stat}}$\,\tablenotemark{a}} &
\multicolumn{2}{c}{This Work} &
\multicolumn{2}{c}{Literature} &
\colhead{References} \\
\cline{4-5} &
\cline{5-6} 
\colhead{} &
\colhead{} &
\colhead{} &
\colhead{$z_{\rm{abs}}$} &
\colhead{{\lognh} [cm$^{-2}$]} &
\colhead{$z_{\rm{abs}}$} &
\colhead{{\lognh} [cm$^{-2}$]} &
\colhead{}
}
\startdata
J0021$-$0128 &  & 1 & 1.2420 & 20.55 $\pm$ 0.10 & 1.2412 & 20.50 & 2\\
J0051$+$0041 &  & 0 & 0.7420 & 20.60 $\pm$ 0.10 & 0.7400 & 20.40 $\pm$ 0.10 & 3\\
J0102$-$0853 &  & 0 & 0.8945 & 20.45 $\pm$ 0.10 & --- & --- & 18\\
J0106$+$0105 &  & 0 & 1.3020 & 20.85 $\pm$ 0.20 & 1.3002 & 20.95 $\pm$ 0.07 & 1\\
J0122$-$2843 & B0120$-$28 & 1 & 0.1856 & 20.55 $\pm$ 0.10 & 0.1856 & 20.50 $\pm$ 0.10 & 4\\
J0126$-$0105 &  & 0 & 1.1930 & 20.65 $\pm$ 0.10 & 1.1916 & 20.60 $\pm$ 0.04 & 1\\
J0139$-$0023 &  & 0 & 0.6840 & 20.60 $\pm$ 0.15 & 0.6828 & 20.60 $\pm$ 0.07 & 1\\
J0153$+$0052 &  & 0 & 1.0610 & 20.35 $\pm$ 0.20 & 1.0599 & 20.43 $\pm$ 0.10 & 1\\
J0304$-$2212 & B0302$-$2223 & 0 & 1.0094 & 20.30 $\pm$ 0.10 & 1.0140 & 20.00 & 5,6\\
J0452$-$1640 &  & 0 & 1.0090 & 21.00 $\pm$ 0.10 & 1.0072 & 20.98 $\pm$ 0.06 & 1\\
J0456$+$0400 & B0454$+$0356 & 0 & 0.8586 & 20.60 $\pm$ 0.10 & 0.8596 & 20.75 $\pm$ 0.02 & 7\\
J0741$+$3111 &  & 0 & 0.2220 & 20.60 $\pm$ 0.20 & 0.2213 & 20.90 $\pm$ 0.07 & 1\\
J0830$+$2410 & B0827$+$2421 & 0 & 0.5191 & 20.40 $\pm$ 0.20 & 0.5247 & 20.30 $\pm$ 0.05 & 1\\
J0930$+$2848 & SDSS-J093001.90$+$284858.4 & 2 & 0.0227 & 20.75 $\pm$ 0.10 & 0.0227 & 20.71 $\pm$ 0.15 & 17\\
J0938$+$4128 & B0935$+$4141 & 1 & 1.3725 & 20.45 $\pm$ 0.15 & 1.3960 & 20.50 & 15\\
J0948$+$4323 &  & 0 & 1.2340 & 21.75 $\pm$ 0.15 & --- & --- & 18\\
J0953$-$0038 &  & 0 & 0.6390 & 20.30 $\pm$ 0.10 & 0.6381 & 19.90 $\pm$ 0.08 & 1\\
J0954$+$1743 &  & 0 & 0.2410 & 21.05 $\pm$ 0.20 & 0.2377 & 21.32 $\pm$ 0.05 & 1\\
J1001$+$5553 & B0957$+$5608a & 0 & 1.3913 & 20.30 $\pm$ 0.20 & 1.3911 & 20.28 $\pm$ 0.08 & 8\\
J1009$+$0713 & SDSSJ100902.06$+$071343.8 & 2 & 0.1139 & 20.75 $\pm$ 0.10 & 0.1140 & 20.68 $\pm$ 0.10 & 9\\
J1009$+$0036 &  & 0 & 0.9730 & 20.30 $\pm$ 0.15 & 0.9714 & 20.00 $\pm$ 0.10 & 1\\
J1010$+$0003 &  & 0 & 1.2670 & 21.70 $\pm$ 0.10 & 1.2651 & 21.52 $\pm$ 0.06 & 1\\
J1017$+$5356 &  & 0 & 1.3070 & 20.70 $\pm$ 0.10 & --- & --- & 18\\
J1106$-$1821 & B1104$-$1805b & 0 & 1.6617 & 20.80 $\pm$ 0.10 & 1.6616 & 20.84 & 10\\
J1107$+$0048 &  & 0 & 0.7410 & 21.00 $\pm$ 0.15 & 0.7470 & 21.00 $\pm$ 0.04 & 1\\
J1124$-$1705 & B1122$-$1648 & 1 & 0.6812 & 20.35 $\pm$ 0.15 & 0.6819 & 20.45 $\pm$ 0.15 & 11\\
J1130$-$1449 & B1127$-$1432 & 0 & 0.3140 & 21.30 $\pm$ 0.15 & 0.3130 & 21.71 $\pm$ 0.07 & 1\\
J1224$+$0037 &  & 0 & 1.2350 & 20.75 $\pm$ 0.15 & 1.2346 & 20.88 $\pm$ 0.05 & 1\\
J1225$+$0035 &  & 0 & 0.7730 & 21.55 $\pm$ 0.10 & 0.7730 & 21.38 $\pm$ 0.12 & 1\\
J1232$-$0224 & Q1232$-$022 & 0 & 0.3950 & 20.85 $\pm$ 0.15 & 0.3950 & 20.75 $\pm$ 0.07 & 12\\
J1251$+$4637 & Q1251$+$463 & 0 & 0.3965 & 20.60 $\pm$ 0.15 & 0.3965 & 20.50 $\pm$ 0.20 & 17\\
J1331$+$3030 & B1328$+$3045 & 0 & 0.6840 & 21.40 $\pm$ 0.15 & 0.6920 & 21.30 & 13\\
J1420$-$0054 &  & 0 & 1.3470 & 20.85 $\pm$ 0.15 & 1.3475 & 20.90 $\pm$ 0.05 & 1\\
J1431$+$3952 &  & 0 & 0.6040 & 21.30 $\pm$ 0.15 & 0.6019 & 21.20 $\pm$ 0.10 & 16\\
J1501$+$0019 &  & 0 & 1.4840 & 20.90 $\pm$ 0.10 & 1.4832 & 20.85 $\pm$ 0.14 & 1\\
J1512$+$0128 & SDSS-J151237.05$+$012846. & 2 & 0.0295 & 20.40 $\pm$ 0.10 & 0.0295 & 20.27 $\pm$ 0.15 & 17\\
J1527$+$2452 &  & 0 & 0.7345 & 20.40 $\pm$ 0.15 & --- & --- & 18\\
J1537$+$0021 &  & 0 & 1.1790 & 20.30 $\pm$ 0.10 & 1.1782 & 20.18 $\pm$ 0.10 & 1\\
J1616$+$4154 & SDSSJ161649.42$+$415416.3 & 2 & 0.3210 & 20.65 $\pm$ 0.20 & 0.3211 & 20.60 $\pm$ 0.20 & 9\\
J1619$+$3342 & SDSSJ161916.54$+$334238.4 & 2 & 0.0964 & 20.65 $\pm$ 0.10 & 0.0963 & 20.55 $\pm$ 0.10 & 9\\
J1624$+$2345 & B1622$+$2352 & 0 & 0.6556 & 20.30 $\pm$ 0.10 & 0.6560 & 20.30 & 14\\
J1712$+$5559 &  & 0 & 1.2100 & 20.65 $\pm$ 0.15 & 1.2093 & 20.72 $\pm$ 0.05 & 1\\
J1727$+$5302 &  & 0 & 0.9480 & 21.25 $\pm$ 0.15 & 0.9448 & 21.16 $\pm$ 0.05 & 1\\
J1727$+$5302 &  & 0 & 1.0330 & 21.50 $\pm$ 0.15 & 1.0312 & 21.41 $\pm$ 0.03 & 1\\
J1733$+$5533 &  & 0 & 0.9990 & 20.80 $\pm$ 0.10 & 0.9981 & 20.70 $\pm$ 0.03 & 1\\
J2334$+$0052 &  & 0 & 0.4740 & 20.50 $\pm$ 0.15 & 0.4713 & 20.65 $\pm$ 0.15 & 1\\
J2339$-$0029 &  & 0 & 0.9680 & 20.60 $\pm$ 0.15 & 0.9664 & 20.48 $\pm$ 0.08 & 1\\
J2353$-$0028 &  & 0 & 0.6044 & 21.50 $\pm$ 0.10 & 0.6044 & 21.54 $\pm$ 0.15 & 1
\enddata
\tablenotetext{a}{DLA is found in the sightline of a quasar that is: (0) not in the statistical sample; (1) in the statistical sample; (2) in the galaxy sample (see Table \ref{tab:qso} and Section \ref{sec:statsample}).}
\tablerefs{(1) \citet{Rao2006}; (2) \citet{Aracil2002}; (3) \citet{Lacy2003}; (4) \citet{Oliveira2014}; (5) \citet{Lanzetta1995}; (6) \citet{Pettini1997}; (7) \citet{Steidel1995}; (8) \citet{Zuo1997}; (9) \citet{Meiring2011}; (10) \citet{Lopez1999}; (11) \citet{Delavarga2000}; (12) \citet{Boisse1998}; (13) \citet{Cohen1994}; (14) \citet{Steidel1997}; (15) \citet{Jannuzi1998}; (16) \citet{Ellison2012}; (17) \citet{Muzahid2015}; (18) This Work.}
\end{deluxetable*}

\subsection{S/N Considerations}
\label{sec:S/N}

Similar to the resolution of a spectrum, the S/N of a spectrum could affect both the
search algorithm's efficacy and ability to accurately measure the
{\hi} column density of the absorption system during the fitting process. In
the case of the search algorithm, low S/N spectra will flag more false
positives as more pixels will satisfy the S/N cut criteria. To prevent high 
rates of false positive detections, we therefore set the minimum S/N ratio to be 4 
over a 20\,{\AA} window, which is 2-$\sigma$ above the S/N cut of 2 per 3\,{\AA} 
utilized in the search algorithm.  

We note that \citet{Noterdaeme2009b} found that DLAs could go undetected in the
search algorithm, if they occur sufficiently blueward in the spectrum as their damping wings 
would keep the S/N below the S/N cut. We have visually checked all of the spectra and find
no evidence that this effect is present in our sample, likely because of the decrease in Ly-$\alpha$
forest density of these low redshift quasar sightlines.

We also need to be able to accurately determine the {\hi} column density of the absorption 
systems with these S/N cuts. To test this, we insert artificial absorption systems 
in the real spectra and increase the noise level to the required S/N values. 
The fitting results are displayed in Figure~\ref{fig:fitdla}. As is clear 
from the figure, we are able to accurately determine the {\hi} column density of systems 
down to S/N$\approx 4$ for even the lowest resolution data in the sample. The dispersion 
of the measured values around the actual {\hi} column density for a spectrum with S/N$ = 4$ 
is 0.15\,dex.

We therefore conclude that a S/N threshold of 4 provides the ideal balance between 
maximizing the search path length while still providing the ability to reliably determine 
the {\hi} column density of all DLAs in the search path. 
\\
\section{Results}
\label{sec:results}

\begin{figure*}[!t]
\epsscale{1.15}
\figurenum{4}
\plottwo{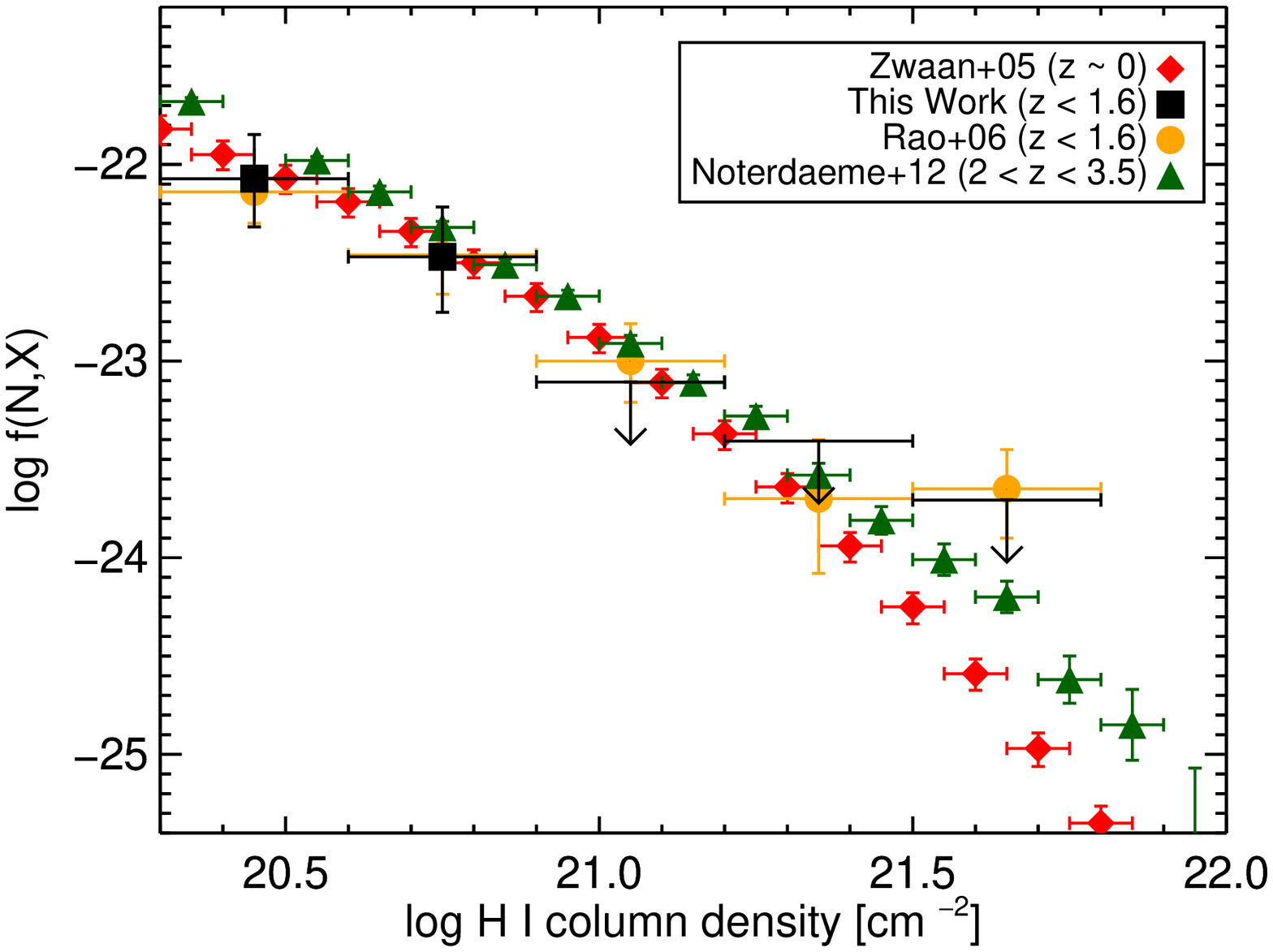}{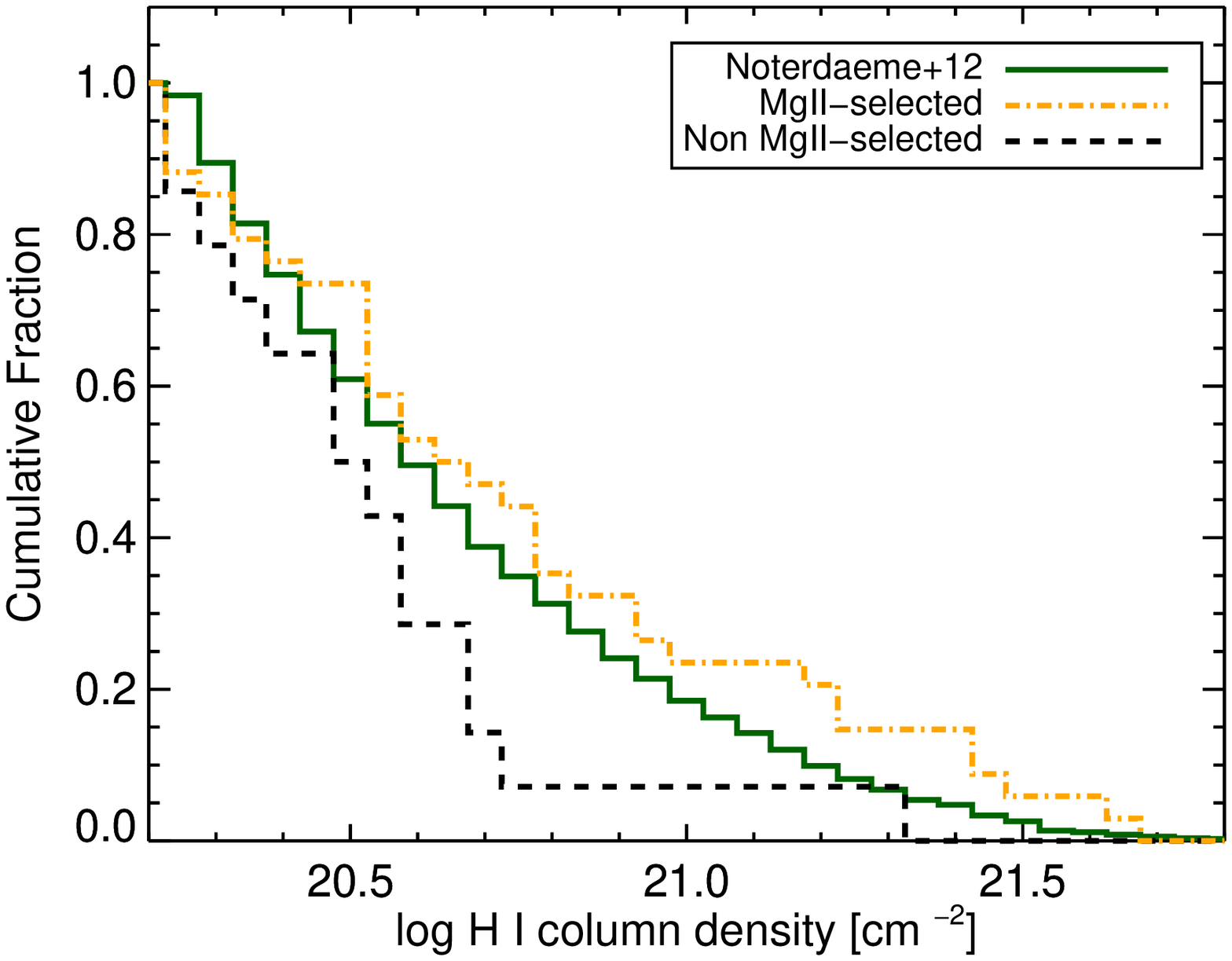}
\caption{The {\hi} column density distribution function, {\fn}, at low redshifts. The black squares
are from the present survey, and described in the text. Our results for {\fn} are consistent with earlier 
{\fn} estimates at both low and high redshifts \citep{Zwaan2005,Noterdaeme2012}. The high value of 
{\fn} found by \citet{Rao2006} at high column densities may be due to a selection effect, which would skew 
the distribution of {\fn} towards higher {\hi} column density. The right panel shows that the cumulative 
distribution of the Mg\,{\sc ii}-selected DLAs is indeed skewed toward higher {\hi} column densities 
(compared to that of the non-Mg\,{\sc ii}-selected DLAs), indicating that this is a plausible selection effect.
However, we emphasize that the sample is not large enough to statistically confirm this 
source of bias.}
\label{fig:fn}
\end{figure*}

\begin{figure}[b]
\epsscale{1.2}
\figurenum{3}
\plotone{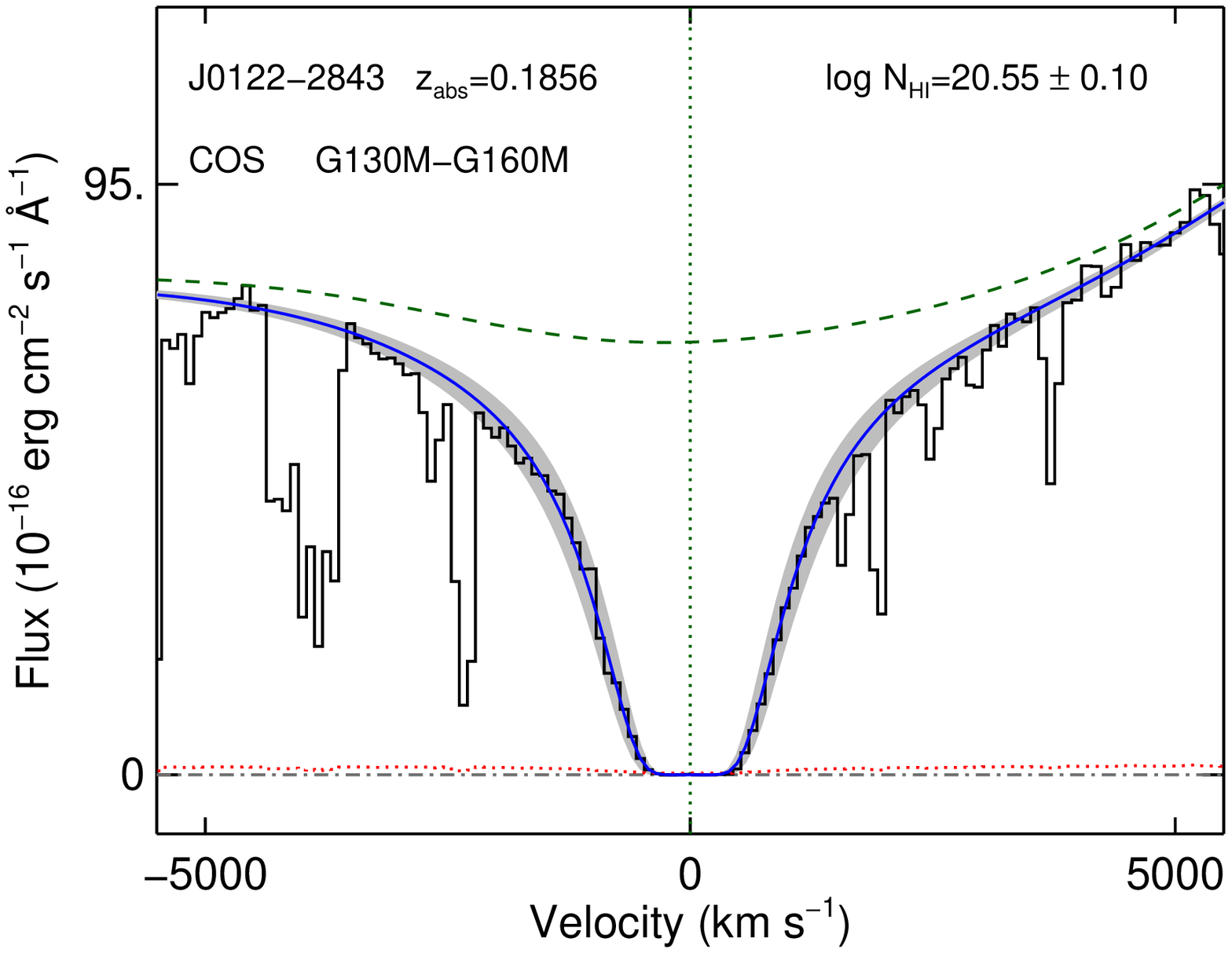}
\caption{Example of the determination of the {\hi} column density of an absorber. 
We simultaneously fit the continuum (dashed green) and the Voigt profile of the 
absorber to determine the {\hi} column density. The gray shaded regions mark the error 
on the fit (solid blue). The dotted red line marks the uncertainty of the data. The full
set of figures for all DLAs listed in Table \ref{tab:dla} is available at the end of this
manuscript or as an online figure set.}
\label{fig:dla}
\end{figure} 

\subsection{The full sample of DLAs} 

Figure~\ref{fig:dla} shows the {\hi} absorption profiles for all the absorbers with a
measured {\hi} column density greater than $10^{20.3}$\,cm$^{-2}$. These
absorbers are also listed in Table \ref{tab:dla}. Our sample contains
a total of 47 DLAs with a mean redshift of $z = 0.796$. Of these 47
systems, 33 were selected due to the presence of strong Mg\,{\sc ii} absorption, 6 
due the presence of a galaxy close to the quasar sightline, 2 were known 
{\hi}~21\,cm absorbers, and 2 were known DLAs. Only \statdla\ DLAs were discovered 
in sightlines not {\it a priori} selected to contain a strong absorber, and 
hence form the statistical sample of this paper. 

We can test our methodology and recovery rate of DLAs by comparing our list of DLAs to those
found in \citet{Rao2006}, as the latter is a subset of this
larger sample. We find that our search algorithm recovers 28 of the 41 DLAs 
in the \citet{Rao2006} sample, with a mean dispersion of 0.10~dex 
between the {\hi} column density estimates. Four of the DLAs were not covered by our
spectra either due to the cut-off at 1350\,{\AA} in the FOS~G160L or some other 
reduction issue. One of these is the serendipitously discovered DLA at z=0.0912 
towards B0738+313. However, we note that the omission of these DLAs do not affect 
the results in this paper, as these sightlines would not have been part of the statistical 
sample because of their pre-selection of Mg\,{\sc ii} absorption. 

Of the remaining 9 DLAs not recovered by our search algorithm, 8 fell in a portion of the spectrum 
with an S/N below our criterion (see Section~\ref{sec:S/N}). We note that these sightlines 
are also not included in our statistical search path and therefore do not affect the conclusions 
presented in this paper. The only DLA (in QSO J0153$+$0052 at $z_{abs} = 1.0599$)
which did fall in the search algorithm's search 
path and was not detected was not found because the trough of this potential DLA 
showed significant flux. This could be either due to Ly-$\alpha$ emission at the 
redshift of the DLA, or an issue with the zero point flux of the spectrum. We have visually 
checked all of the spectra and believe that this is a very rare occurrence. 

\subsection{The Column Density Distribution Function, {\fn}}
\label{sec:fn}

We obtain the {\hi} column density distribution function with the approach described 
in, e.g., \citet{Tytler1987}. {\fn} $dN_{\rm{H}\textsc{i}}\,dX$ is defined as the number of 
DLAs with \hi\ column density between $N_{\rm{H}\textsc{i}}$ and $N_{\rm{H}\textsc{i}I} + 
dN_{\rm{H}\textsc{i}}$ and within the absorption distance $dX$. $dX$ is defined as:
\begin{equation}
dX \equiv \frac{H_0}{H(z)} (1+z)^2 dz,
\end{equation}
where $H_0$ is the Hubble constant and $H(z)$ is given by:
\begin{equation}
H(z) = H_0 [(1+z)^2(1+z\Omega_m)-z(z+2)\Omega_{\Lambda}]^{-1/2}.
\end{equation}
The absorption distance is defined in this manner such that {\fn} is constant for 
a non-evolving population of absorbers. 

\begin{figure*}[t!]
\epsscale{1.15}
\figurenum{5}
\plottwo{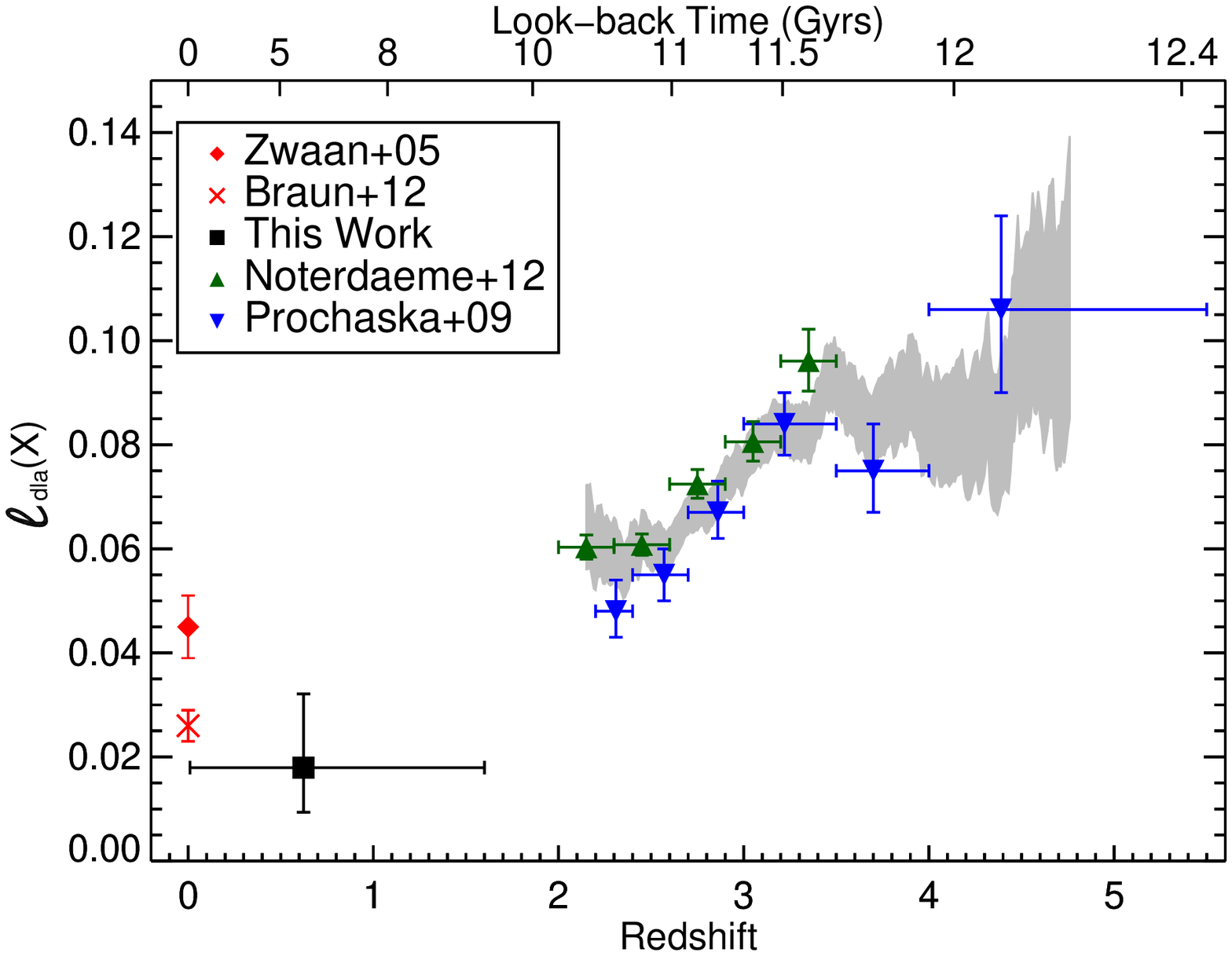}{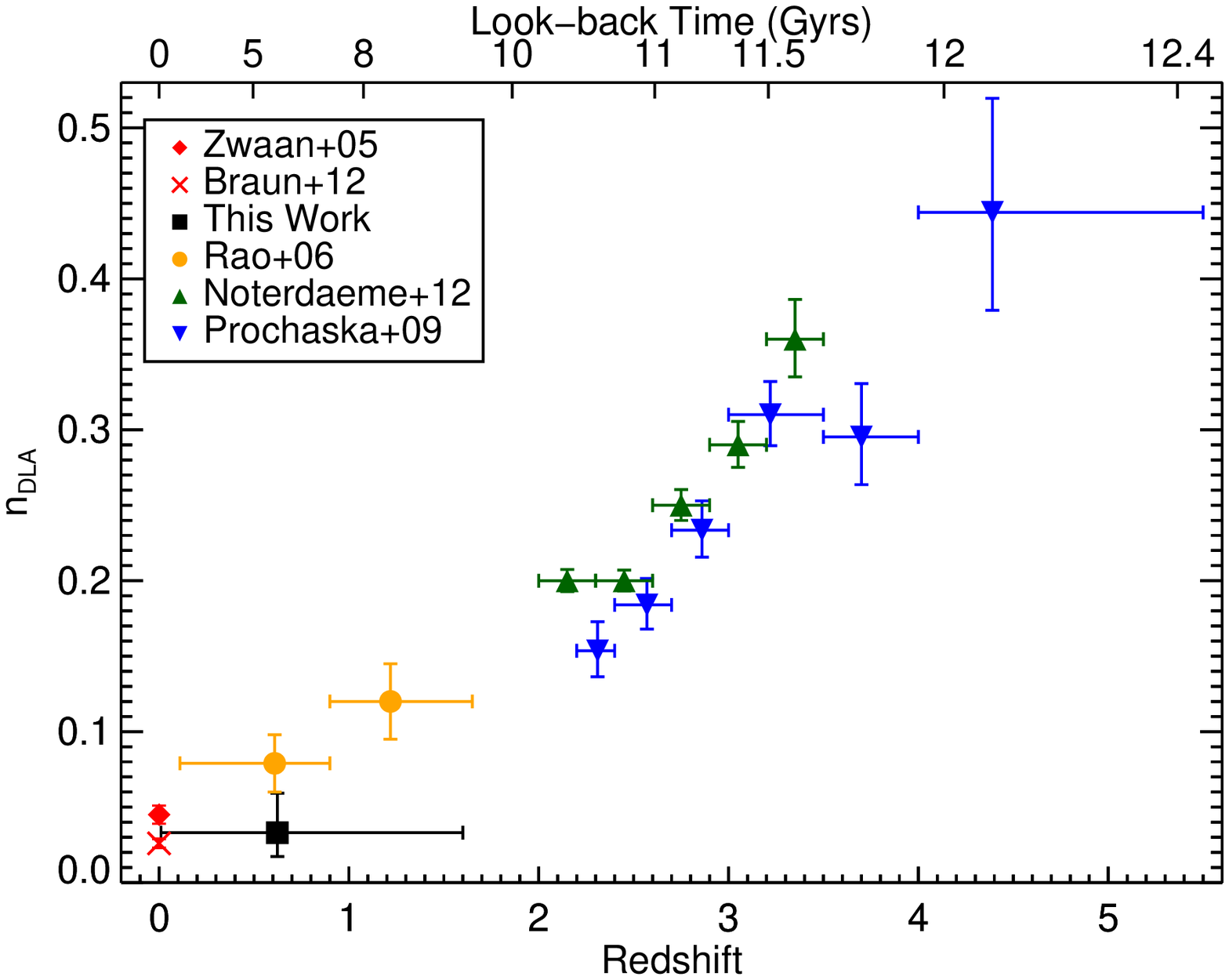}
\caption{\emph{Left panel:} Line density of DLAs, {\lox} and \emph{Right panel:} 
Redshift number density of DLAs, $n_{\rm{DLA}}$. The gray area marks the 68\,\% 
confidence level for the sample compiled by \citet{Sanchez2015}. The line 
density of DLAs over the redshift interval $z \approx 0.01$ to $z \approx 1.6$ is 
lower than the measured line density at redshift $z \approx 2$ at greater than
2-$\sigma$ significance, indicating that {\lox} has evolved over the past 10 Gyrs.}
\label{fig:lox}
\end{figure*} 

The statistical sample only contains \statdla\ DLAs, and therefore we
have poor constraints on the functional form of {\fn}. To increase the
sample size, we include all sightlines selected to probe an
intervening galaxy (i.e. those with f$_{\rm{stat}}$ = 2 in Table
\ref{tab:qso}), resulting in a DLA sample of 9 DLAs. We note that the inclusion 
of these sight lines will
likely bias the normalization of {\fn} high. The resulting {\fn} is
shown in the left panel of Figure~\ref{fig:fn}. Our results are consistent with the
results of the local {\hi}~21\,cm study of \citet{Zwaan2005}, and 
with those of DLA surveys at higher redshifts \citep[e.g.,][]{Prochaska2009, Noterdaeme2012}. 
However, our sample size remains too small to constrain {\fn} at the highest 
{\hi} column densities. 

We also consider {\fn} for the sample described in
\citet{Rao2006}. They note that their sample contains a large number
of high {\hi} column density systems, although this deviation is
not statistically significant. We have broken up our sample into those
DLAs that were found using the Mg\,{\sc ii} selection criteria defined
in \citet{Rao2006} and the remaining DLAs. The cumulative distribution
plotted in the right panel of Figure~\ref{fig:fn} indeed shows that the
Mg\,{\sc ii}-selected DLA sample has a larger number of high {\hi} column density
systems than the rest of the low-$z$ DLA sample (this work) or the high-$z$ DLA sample 
of \citet{Noterdaeme2012}. This corroborates the suggestion of \citet{Prochaska2009} 
that selecting sightlines based on metal line absorption could bias the {\hi} distribution 
toward high column densities. This result is also seen in recent work on metal-strong
absorbers \citep{Dessauges2009, Kaplan2011, Berg2015}. 

\subsection{Line density of DLAs, {\lox}}
\label{sec:lox}

The line density of DLAs, {\lox}, is defined as the zeroth moment of {\fn}, i.e.:
\begin{equation}
 \ell_{\rm{DLA}}(X) = \int^{\infty}_{N_{\rm{DLA}}} f(N_{\rm{H{\scriptscriptstyle\,I}}},X) dNdX,
\end{equation}
where $N_{\rm{DLA}}$ is the threshold {\hi} column density for
DLAs. In practice, {\lox} is estimated by measuring the number of DLAs
in a given redshift bin and then dividing this value by the total absorption path
length in this redshift bin. A correlated quantity is the redshift number density, 
$n_{\rm{DLA}}$, which is obtained by dividing the number of DLAs found in a
redshift bin by the redshift path length. Both of these quantities describe the 
incidence rate of DLAs along a line of sight.  

We have calculated {\lox} and $n_{\rm{DLA}}$ for the complete statistical sample
(those with f$_{\rm{stat}} = 1)$, which covers the redshift range, $z = 0.01 - 1.6$, 
with a median redshift of $0.623$. The resultant {\lox} is $0.017^{+0.014}_{-0.008}$, 
which equates to $n_{\rm{DLA}}$ = $0.033^{+0.026}_{-0.015}$. These values are put in context 
by plotting them with a compilation from the literature in Figure~\ref{fig:lox}. 

Figure~\ref{fig:lox} shows that the line density obtained from this sample is lower 
than the locally measured values of \citet{Zwaan2005} and \citet{Braun2012} (note that
 the latter estimate is based on an extrapolation from just three galaxies), although our 
estimate is consistent with these values within 2-$\sigma$ significance. Our {\lox} estimate 
is also lower than the estimate of \citet{Rao2006}, but the systematic uncertainties in their 
measurement are difficult to quantify.

Our new estimate of the line density of DLAs at $z < 1.6$ is lower than
this value at $z \sim 2$ \citep{Noterdaeme2012} at  greater than 2-$\sigma$ significance. 
This result suggests that {\lox} has evolved over this redshift range. This result is 
discussed further in Section~\ref{sec:disc}.

We note that we have searched for evolution within our sample by dividing 
the sample into two subsamples at the median redshift of $z=0.623$. However, the sample
size is too small to discern any evolution within the sample, although the sample is 
consistent with an evolution of $< 0.05$ in {\lox} per unit redshift, which agrees with the evolution
at higher redshift ($\sim 0.03$ in {\lox} per unit redshift, see e.g., \citet{Sanchez2015}).

\subsection{Mass Density of {\hi}, {\rhohi}}

The final quantity we consider is the first moment of {\fn}, which is the mass 
density of {\hi} contained in DLAs, {\rhohi}. This quantity is defined by:
\begin{equation}
\rho_{\rm{H{\scriptscriptstyle\,I}}}^{\rm{DLA}} = \frac{m_{\rm{H}}H_0}{c} \int^{N_{\rm{max}}}_{N_{\rm{min}}} N_{\rm{H{\scriptscriptstyle\,I}}} f(N_{\rm{H{\scriptscriptstyle\,I}}},X) dNdX,
\end{equation}
where $m_{\rm{H}}$ is the mass of the hydrogen atom and $c$ is the speed of light. 
The superscript clarifies that we are only measuring the fraction of 
{\hi} in DLAs; of course, as noted earlier, these systems contain the bulk ($> 85$\,\%) of 
the {\hi} at all redshifts \citep{Zwaan2005,OMeara2007,Noterdaeme2012}. We also note
that literature studies often provide estimates of the cosmological neutral gas mass density, 
$\Omega_{g}^{\rm{DLA}}$. {\rhohi} is related to $\Omega_{g}^{\rm{DLA}}$ by the conversion 
factor $\mu/\rho_c$, where $\mu$ is the mean molecular mass of the gas and $\rho_c$ is 
the critical density of the Universe.

As with {\fn}, {\rhohi} cannot be accurately determined from our own study because the small sample 
size lacks the statistics to accurately determine the number of high {\hi} column density (\nh~$\gtrsim 10^{21}$~cm$^{-2}$) 
systems (none are present in our survey). These systems are likely to contribute significantly to {\rhohi} 
\citep{Zwaan2005, OMeara2007, Noterdaeme2012}, and our results are hence likely to underestimate the underlying neutral 
gas density. To account for this, we assume that the mean column density of the low redshift absorber sample 
is unchanged from the mean column density at high redshifts, as measured by \citet{Noterdaeme2012} 
(see Section \ref{sec:fn}).

\begin{figure}[t!]
\epsscale{1.15}
\figurenum{6}
\plotone{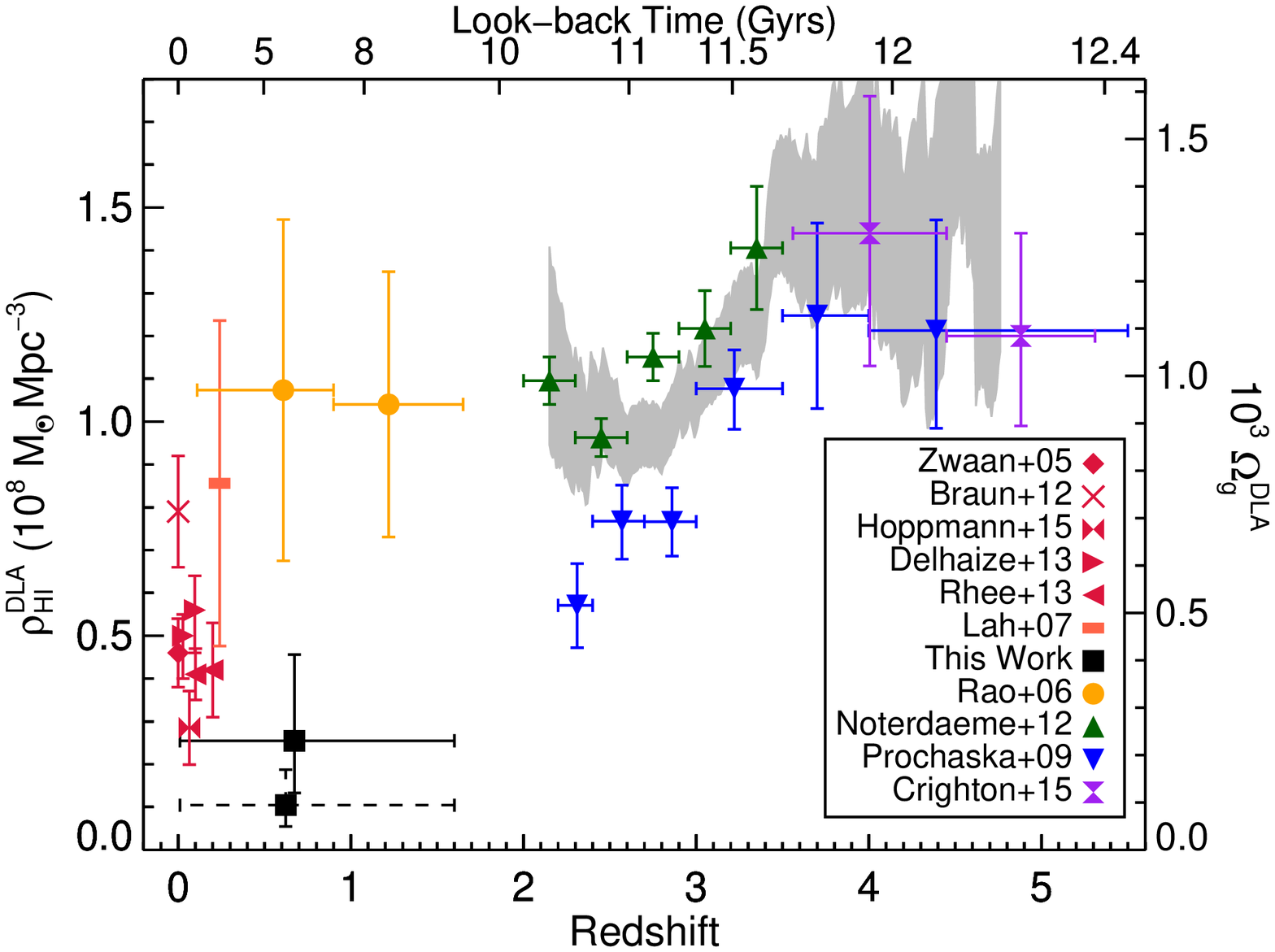}
\caption{Mass density of {\hi} in DLAs, {\rhohi}. The dashed point is the 
unadjusted measurement not accounting for the small sample size, whereas
the solid black point accounts for the sample size (see text). The gray area, 
as in Figure \ref{fig:lox}a, marks the 68\,\% confidence level of {\rhohi} for the 
sample compiled by \citet{Sanchez2015}. The literature
estimates of $\Omega_{\rm{H{\scriptscriptstyle\,I}}}$ were 
converted to {\rhohi} by assuming 85\,\% of the cosmic {\hi} is in DLA gas.
The data indicates a slow decrease in {\rhohi} over the past 10 Gyrs.} 
\label{fig:rhohi}
\end{figure} 

Our data yield {\rhohi}=($0.10^{+0.08}_{-0.05} \times 10^8 M_{\odot} \rm{Mpc}^{-3}$) 
over the redshift range $0.01 \lesssim z \lesssim 1.6$, without any correction for the 
missing high \hi\ column density absorbers. Using the mean \hi\ column density estimate 
of \citet{Noterdaeme2012} to account for their absence, we obtain a corrected 
{\rhohi} of $0.25^{+0.20}_{-0.12} \times 10^8 M_{\odot} \rm{Mpc}^{-3}$.
Both the corrected (solid error bars) and uncorrected (dashed error bars) values are 
shown in Figure~\ref{fig:rhohi}, along with other estimates of this quantity at different 
redshifts from the literature.

Our corrected estimate of {\rhohi} is lower than the earlier estimate over the same 
redshift range by \citet{Rao2006}. The lower value compared to that of \citet{Rao2006} is due 
to both the decrease in {\lox} and the lower adopted mean {\hi} column density (see Section~\ref{sec:fn}). 
Our {\hi} mass density estimate is, however, in good agreement with the estimates from {\hi}~21\,cm studies 
at $z \approx 0-0.3$ \citep{Zwaan2005,Lah2007,Delhaize2013,Rhee2013,Hoppmann2015}. The results suggest a mild evolution
 in {\rhohi}, driven by the evolution in {\lox}, which was recently quantified by 
\citet{Crighton2015} and \citet{Sanchez2015} who measured {\rhohi} at high redshifts.
 
Finally, we would like to point out the apparent discrepancy in the {\rhohi} measurements
of \citet{Prochaska2009} and \citet{Noterdaeme2012} at $z \approx 3$. This discrepancy is 
due to two factors: (i)~the sample analyzed by \citet{Prochaska2009} led to an under-estimate in 
the break in {\fn} by $\approx 0.2$\,dex and therefore an under-estimate of $\rho_{\rm HI}$ 
by $\approx 20$\,\%; (ii)~the results are now limited by systematic uncertainties that include 
the approach to measuring {\nh}, dust obscuration, and color-selection bias in quasar surveys. 
These systematic errors must be addressed before further progress can be made in this field.

\section{Summary and Discussion}
\label{sec:disc}

In this paper, we have used a large sample of quasars ($\nqso$ systems)
observed with the UV spectrographs on the Hubble Space Telescope to
carry out a search for low-redshift DLAs. Of the 46 DLAs found in this study, 
only \statdla\ are drawn from the statistical sample. The remainder were drawn 
from sightlines with foreknowledge of either an absorber or an intervening galaxy,
or the lack of an absorber, close to or along the quasar line of sight. 

Our statistical sample enables an unbiased determination of the line density of 
DLAs, {\lox}, over the redshift range $0.01 \lesssim z \lesssim 1.6$. We obtain 
{\lox}~$= 0.017^{+0.014}_{-0.008}$, significantly smaller than the previous estimate of 
\citet{Rao2006}, which appears likely to be biased high. Unfortunately, current estimates 
of {\lox} at these redshifts continue to suffer from the small sample size of quasar sightlines.

Previous studies, e.g., \citet{Prochaska2005,Rao2006,Prochaska2009,Noterdaeme2012}, 
have claimed little evolution in the line density of DLAs in the past 10\,Gyrs. 
However, the results in this paper indicate that the line density at $z \approx 0.5$ is 
lower than estimates of this quantity at $z \approx 2$ at greater than 2-$\sigma$ significance.
This suggests a mild evolution in {\lox} from $z \approx 2$ to the present epoch, instead of the inferred 
constancy in the DLA line density. The decrease in {\lox} at low redshifts can be explained 
if the majority of galactic-scale dark matter halos are fully assembled by $z \approx 2$, and, 
if the neutral hydrogen content of these halos slowly decreases, either by star formation or 
feedback processes.

One caveat to this result is the possibility of systematic errors that can bias our estimates
low. Here we discuss three of these biases. One potential bias to our sample is the exclusion of 
sightlines that were targeted to pass close to foreground galaxies. A random sample of quasars 
would contain some sightlines that pass by intervening galaxies, and by removing all of these 
quasars from the sample, we could be biasing our result low. To estimate the effect of this 
potential bias, we include all sightlines that are specifically chosen to cross a foreground 
galaxy (i.e. the quasars with f$_{\rm{stat}}$ = 2). The resultant line density is {\lox}~$=0.030^{+0.014}_{-0.010}$, 
which is well within the 1-$\sigma$ statistical uncertainty of our original estimate, indicating 
that this bias is smaller than our uncertainties. 

A second possible bias may arise due to the inclusion of sightlines at redshifts $z \gtrsim 0.3$.
For these redshifts, metal lines (in particular Mg\,{\sc ii}) would fall within the optical part 
of the quasar spectrum. The target selection criteria for the quasar sample may have included a 
lack of metal line systems in the optical part of the quasar spectrum. Similarly, by excluding 
all of the sightlines with known Mg\,{\sc ii} systems, we might be biasing ourselves against quasars 
with metal lines in their sightlines. To test the bias in our sample against Mg\,{\sc ii} systems, we 
have compared the line density of Mg\,{\sc ii} systems in our quasar sample with the line density in 
the sample of \citet{Seyffert2013}. We find that the line density of Mg\,{\sc ii} systems is 0.119 over the 
full redshift range probed by the quasars in our statistical sample, which are also part of the sample 
of \citet{Seyffert2013} (n=93 quasars). This value agrees well with the measured line density of 
$0.123 \pm 0.001$ over the slightly larger redshift range covered by the full sample of \citet{Seyffert2013}, 
and hence we assert that this effect is unlikely to significantly bias our results.

Finally, the third possible source of bias is that against sightlines with a large amount 
of dust. This bias has been studied in detail using spectroscopy of radio-selected 
samples in high-$z$ DLAs \citep[e.g.,][]{Ellison2001,Akerman2005,Jorgenson2006}
as well as dust reddening of quasars in SDSS \citep[e.g.,][]{Murphy2004,Frank2010,
Fukugita2015,Murphy2016}. This bias is correlated with the amount of metals in 
the gas \citep{Vladilo2005}, and therefore is amplified at low-$z$ due to the higher metallicity 
of low-$z$ DLAs \citep[e.g.,][]{Prochaska2003,Rafelski2012,Rafelski2014}.

Following \citet{Vladilo2005}, dust extinction becomes significant at zinc column densities greater 
than $\sim 10^{13.5}$\,cm$^{-2}$, an assertion corroborated by the few DLAs showing distinct
dust signatures \citep{Noterdaeme2009a,Kulkarni2011,Ma2015}. For low column density DLAs 
(log {\nh} $<$ 20.5) these zinc column 
densities imply $[M/H] >0.5$. Such systems are expected to be rare, both from metallicity measurements 
of lower column density systems, and metallicity measurements from star-forming galaxies 
\citep[e.g.][]{Tremonti2004}. Since {\lox} is dominated by low {\hi} column density systems, we assert that 
the effect of dust on biasing this result is small. In passing, we note that the amount of dust reddening in 
quasars with DLAs is observed to be small at $z \approx 2$ \citep{Murphy2004}, and evolves only slightly 
between redshifts 2.1 and 4.0 \citep{Murphy2016} further suggesting that dust is not likely a large
biasing factor.

We estimate the {\hi} column density distribution function, {\fn}, and the {\hi} gas density, 
{\rhohi} from our statistical sample. However, the limited sample size means that the errors on the inferred 
quantities are large. We find that the {\hi} column density distribution function in our low redshift sample 
is consistent with the {\fn} measured at high redshift. The increase in the high {\hi} column density systems 
found by \citet{Rao2006} could be due to their selection process, but we cannot establish this assertion 
with the current dataset. 

The {\hi} mass density, {\rhohi}, is estimated to be $0.25^{+0.20}_{-0.12} \times 10^8 M_{\odot} \rm{Mpc}^{-3}$,
after correcting for the lack of statistics at the high {\hi} column density end due to our
small sample size of DLAs. The correction makes the assumption that the mean {\hi} column density 
of our low-$z$ DLA sample is the same as that of the high-$z$ DLA sample, as measured 
from the SDSS surveys \citep{Noterdaeme2012}. Our value of {\rhohi} is markedly lower than that obtained 
in previous measurements at these redshifts, but is in agreement with estimates of the {\hi}
mass density at $z \lesssim 0.3$ using {\hi}~21\,cm studies 
\citep{Zwaan2005,Lah2007,Delhaize2013,Rhee2013,Hoppmann2015}. 
The apparent mild evolution seen in {\rhohi} from $z \approx 2$ arises due to the evolution in 
{\lox}, since we have assumed a non-evolving mean {\hi} column density.

In conclusion, this study has compiled the largest sample of quasars observed in the UV 
with spectral resolution sufficient to search for low-redshift DLAs. Even with a total 
of {\nqso} quasars in our target sample, the final sample of DLAs in these
spectra is small, containing only 46 DLAs. This is in part due to the smaller than 
expected incidence rate of DLAs compared to the measurements from Mg\,{\sc ii} surveys 
\citep{Rao2006}. However, the lower incidence rate is in agreement with both the results 
at $z \approx 2$ and $z \approx 0$, if we assume a mild decrease in {\lox} over this redshift range.\\
\\
This research would not have been possible without the guidance and insights of the 
late A.~M.~Wolfe during the initial phases of the project. He will be sorely missed. 
We thank R. S\'{a}nchez-Ram\'{i}rez for providing their results before publication, and the
referee for helpful comments that improved the manuscript.
This work was based on observations made with the NASA/ESA Hubble Space Telescope, 
obtained from the data archive at the Space Telescope Science Institute (STScI). 
The COS G130M/G160M data presented in this work were obtained from the COS-CGM Legacy 
database, which is funded by NASA through grant HST-AR-12854 from the STScI. 
Finally, support for this work was provided by NASA through grant HST-AR-12645
from the STScI. STScI is operated by the Association of 
Universities for Research in Astronomy, Inc. under NASA contract NAS 5-26555. 
MN and JXP further acknowledge support from NSF award AST-1109452.
NL acknowledges support provided by NASA through grant HST-AR-12854. JCH and
NL acknowledge support from NSF award AST-1212012. MR acknowledges support 
from an appointment to the NASA Postdoctoral Program at Goddard Space Flight Center.
NK acknowledges support from the Department of Science and Technology via a 
Swarnajayanti Fellowship (DST/SJF/PSA-01/2012-13). 

\bibliography{./bib}

\clearpage
\begin{figure*}[t]
\epsscale{1.}
\renewcommand\thefigure{3}
\plotone{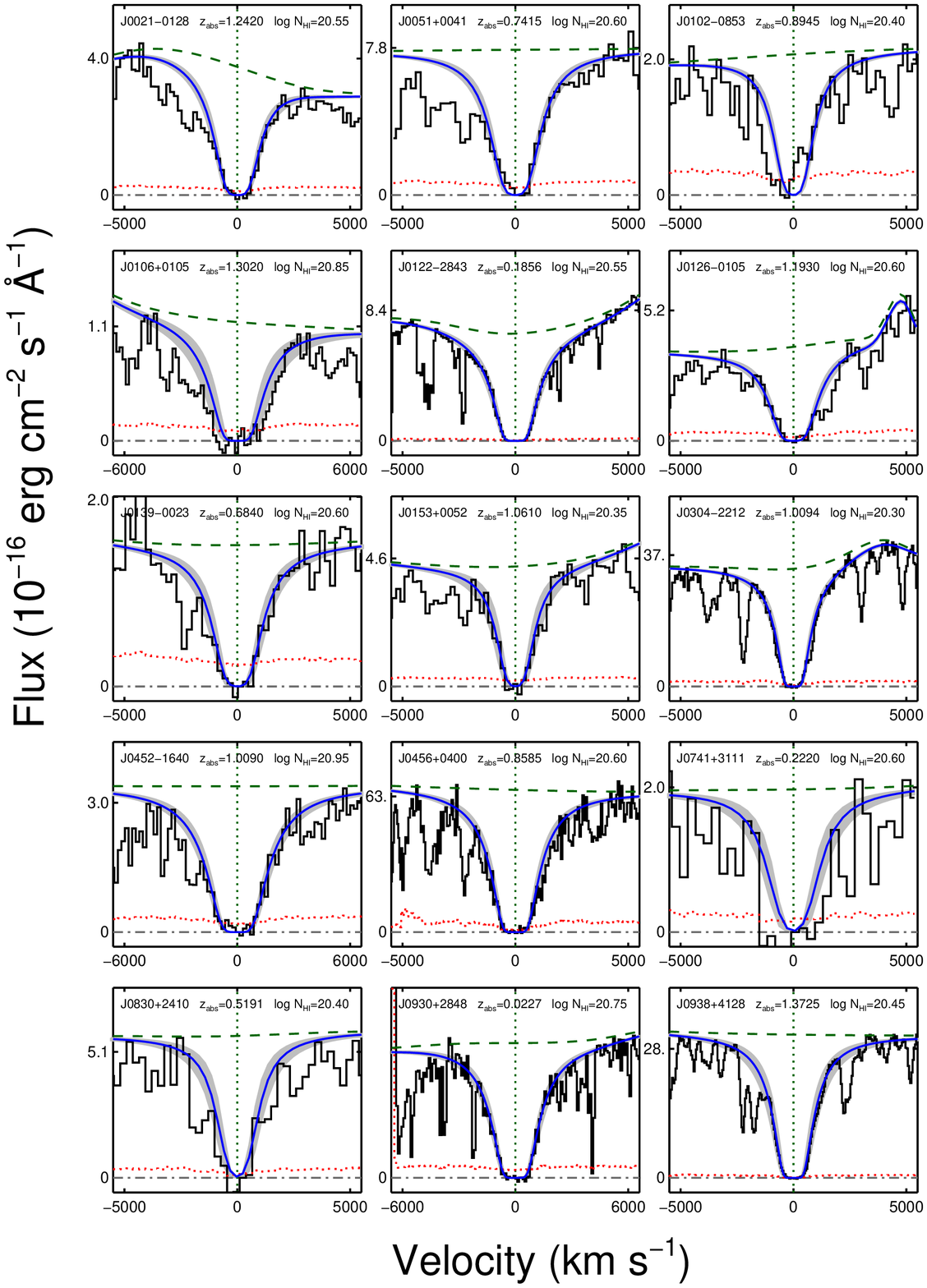}
\caption{$-$ Continued}
\label{fig:dla1}
\end{figure*} 

\begin{figure*}[t]
\epsscale{1.}
\renewcommand\thefigure{3}
\plotone{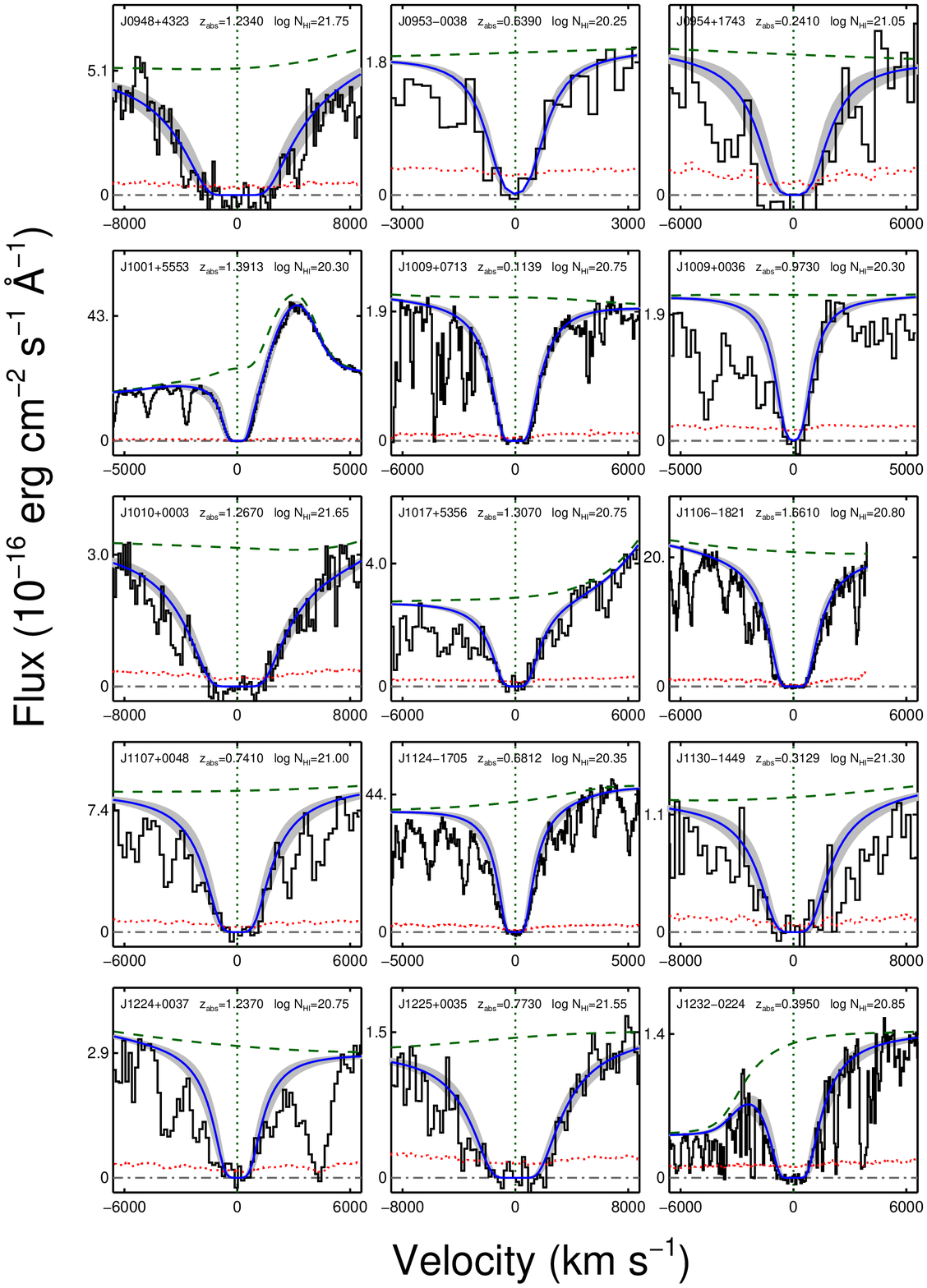}
\caption{$-$ Continued}
\label{fig:dla2}
\end{figure*} 

\begin{figure*}[t]
\epsscale{1.}
\renewcommand\thefigure{3}
\plotone{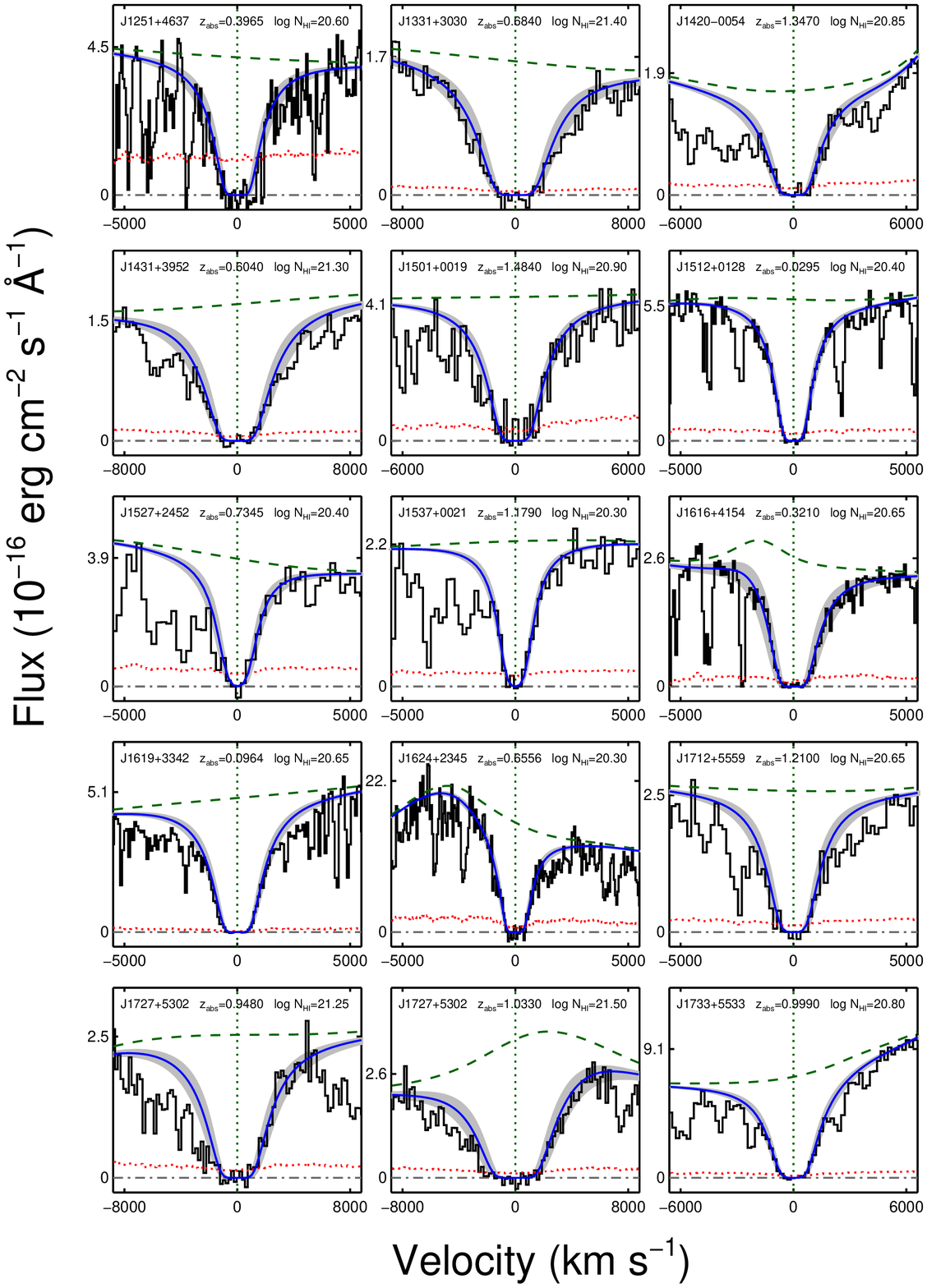}
\caption{$-$ Continued}
\label{fig:dla3}
\end{figure*} 

\begin{figure*}[t]
\epsscale{1.}
\renewcommand\thefigure{3}
\plotone{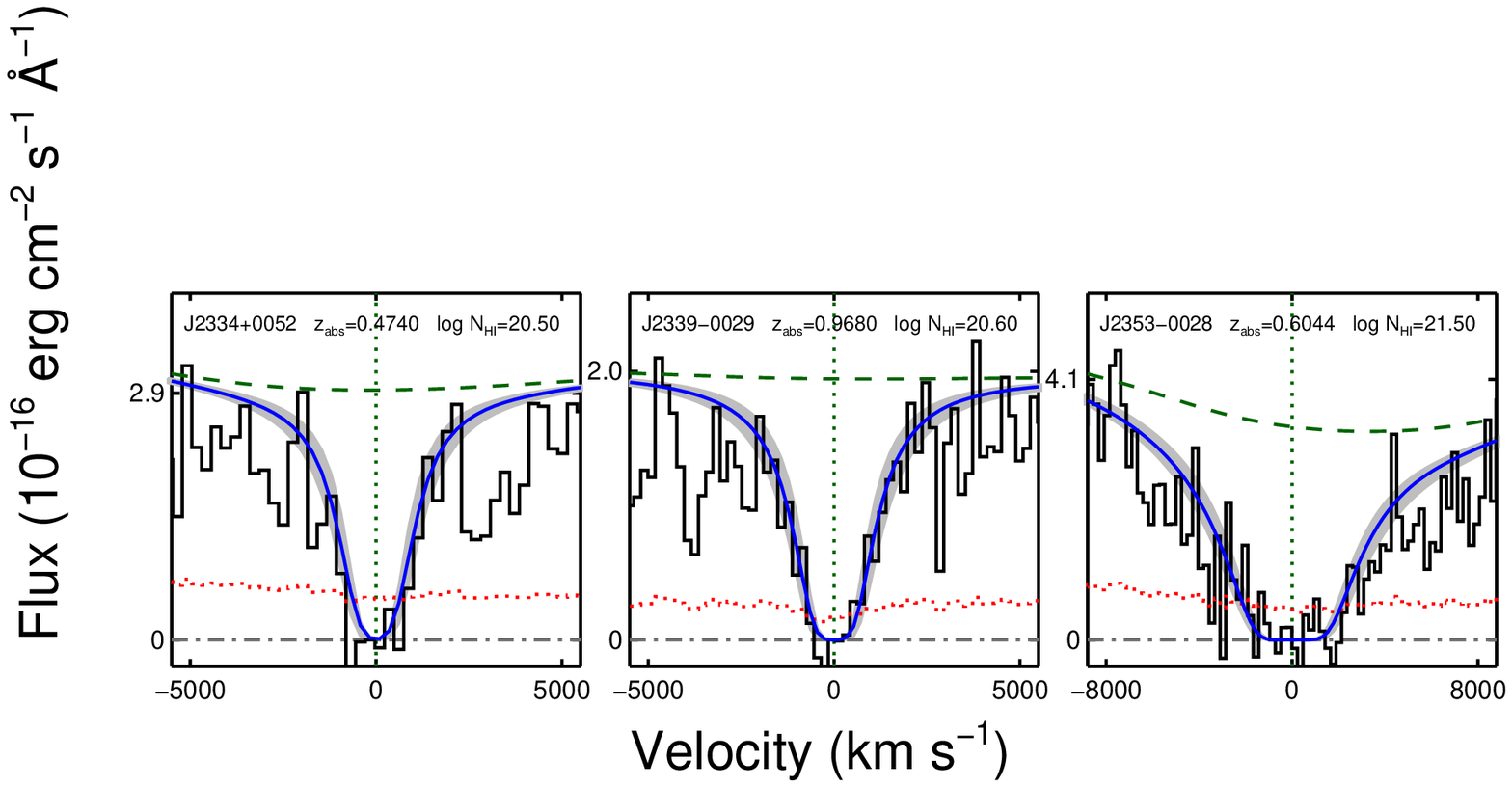}
\caption{$-$ Continued}
\label{fig:dla4}
\end{figure*} 

\clearpage
\LongTables
\begin{turnpage}
\begin{deluxetable*}{llllcllclll}
\tabletypesize{\scriptsize}
\tablecaption{{QSO sample}
\label{tab:qso_long}}
\tablenum{1}
\tablewidth{0pt}
\tablehead{
\colhead{QSO} &
\colhead{$z_{\rm{em}}$\,\tablenotemark{a}} &
\colhead{Instrument} &
\colhead{Grating} &
\multicolumn{3}{c}{Search Path} &
\multicolumn{3}{c}{Statistical Path} &
\colhead{Proposal ID} \\
\cline{5-7} &
\cline{7-9} 
\colhead{} &
\colhead{} &
\colhead{} &
\colhead{} &
\colhead{F$_{\rm{search}}$\,\tablenotemark{b}} &
\colhead{Min $z$} &
\colhead{Max $z$} &
\colhead{F$_{\rm{stat}}$\,\tablenotemark{c}} &
\colhead{Min $z$} &
\colhead{Max $z$} &
\colhead{}
}
\startdata
J0000$-$1245 & 0.200 & COS & G130M-G160M & 1 & 0.010 & 0.299 & 1 & 0.010 & 0.188 & 12604\\
J0001$+$0709 & 3.234 & STIS & G230L & 1 & --- & --- & 0 & --- & --- & 8569\\
J0004$-$4157 & 2.760 & FOS & G270H & 1 & 1.501 & 1.695 & 0 & --- & --- & 6577\\
J0005$+$0524 & 1.900 & FOS & G270H-G190H & 1 & 0.829 & 1.695 & 1 & 0.829 & 1.695 & 4581,6705\\
J0005$-$5006 & 0.033 & COS & G130M-G160M & 1 & 0.010 & 0.133 & 1 & 0.010 & 0.022 & 12936\\
J0005$+$0202 & 0.234 & FOS & G190H & 1 & --- & --- & 1 & --- & --- & 5456\\
J0005$+$1609 & 0.450 & COS & G130M-G160M & 1 & 0.010 & 0.474 & 1 & 0.010 & 0.436 & 12038\\
 &  & FOS & G190H-G270H-G130H & 1 & 0.475 & 0.550 & 1 & --- & --- & 3791\\
\enddata
\tablenotetext{a}{Emission redshift of quasar.}
\tablenotetext{b}{Search flag: (0) Low S/N or bad spectrum; (1) Included; (2) BAL quasar.}
\tablenotetext{c}{Statistical flag: (0) Non-Statistical; (1) Statistical; (2) Galaxy Sample.}
\end{deluxetable*}

\clearpage
\end{turnpage}
\end{document}